\documentclass[11pt]{article}
\usepackage[dvips]{color}
\usepackage{array}
\usepackage{epsf} 
\usepackage{amsmath,color}
\usepackage{amssymb}
\usepackage{epsfig}
\usepackage{latexsym}
\usepackage{amsfonts}
\usepackage{dcolumn}
\usepackage{varioref}
\usepackage{ifthen}
\usepackage{graphicx}
\usepackage{amssymb,amsmath,amsthm}
\usepackage{slashed}
\begin{document}
\parindent 0mm 
\setlength{\parskip}{\baselineskip} 
\thispagestyle{empty}
\pagenumbering{arabic} 
\setcounter{page}{1}
\begin{center}
	{\Large {\bf Finite Temperature QCD Sum Rules: a Review}}\\
\end{center}	
\begin{center}
{\bf Alejandro Ayala}$^{(a),(b)}$,
{\bf  C. A. Dominguez}$^{(b)}$,
{\bf M. Loewe}$^{(b),(c), (d)}$
\end{center}

\begin{center}
{\it $^{(a)}$ Instituto de Ciencias Nucleares, Universidad Nacional Autonoma de Mexico, Apartado Postal 70-543, Mexico, D.F. 04510, Mexico}\\
	
{\it $^{(b)}$Centre for Theoretical and Mathematical Physics, 
	and Department of Physics, University of
	Cape Town, Rondebosch 7700, South Africa}\\

{\it $^{(c)}$Instituto de Fisica, Pontificia Universidad Catolica de Chile, Casilla 306, Santiago 22, Chile}\\

{\it ${(d)}$ Centro Cientıfico-Tecnologico de Valparaiso, Casilla 110-V, Valparaiso, Chile}
\end{center}

\begin{abstract}
\noindent	
 The method of QCD sum rules at finite temperature is reviewed, with emphasis on recent results. These include predictions for the survival of charmonium and bottonium states, at and beyond the critical temperature for de-confinement, as later confirmed by lattice QCD simulations. Also included are determinations in the light-quark vector and axial-vector channels, allowing to analyse the Weinberg sum rules, and predict the dimuon spectrum in heavy ion collisions in the region of the rho-meson. Also in this sector, the determination of the temperature behaviour of the up-down quark mass, together with the pion decay constant, will be described. Finally, an extension of the QCD sum rule method to incorporate finite baryon chemical potential is reviewed.
\end{abstract}

\noindent
\section{Introduction}
The purpose of this article is to review  progress over the past few years on the thermal behaviour of hadronic and QCD matter obtained within the framework of QCD sum rules (QCDSR) \cite{SVZ}-\cite{Review1} extended to finite temperature, $T \neq 0$. These thermal QCDSR were first proposed long ago by Bochkarev and  Shaposnikov \cite{BS}, leading to countless applications, with the most recent ones being reviewed here. The first step in the thermal QCDSR approach is to identify the relevant quantities to provide information on the basic phase transitions (or crossover), i.e. quark-gluon de-confinement and chiral-symmetry restoration. This is done below, to be followed in Section 2 by a brief description of the QCD sum rule method at $T=0$, which relates QCD to hadronic physics by invoking Cauchy's theorem in the complex square-energy plane. Next, in Section 3 the extension to finite $T$ will be outlined in the light-quark axial-vector channel, leading to an intimate relation between de-confinement and chiral-symmetry restoration. In Section 4 the thermal  light-quark  vector channel is described, with an application to  the di-muon production rate in heavy-ion collisions at high energies, which can be predicted in the $\rho$-meson region in excellent agreement with data. Section 5 is devoted to the thermal behaviour of the Weinberg sum rules, and the issue of chiral-mixing. In Section 6 very recent results on the thermal behaviour of the up- down-quark mass will be shown.  Section 7 is devoted to the thermal behaviour of heavy-quark systems, i.e.  charmonium and bottonium states, which led to the prediction of their survival at and above the critical temperature for de-confinement, confirmed  by lattice QCD (LQCD) results. In Section 8 we review an extension of the thermal QCDSR method to finite baryon chemical potential.
Finally, Section 9 provides a short summary of this review.\\
\noindent
Figure 1 illustrates a typical hadronic spectral function, ${\mbox{Im}} \,\Pi(s)$, in terms of the square energy, $s\equiv E^2$, in the time-like region, $s>0$, at $T=0$ (curve (a)). First, there could be a delta-function corresponding to a stable particle present as a pole (zero-width state) in the spectral function. 
\begin{figure}[ht]
	\begin{center}
		\includegraphics[height=3.2in, width=4.8in]{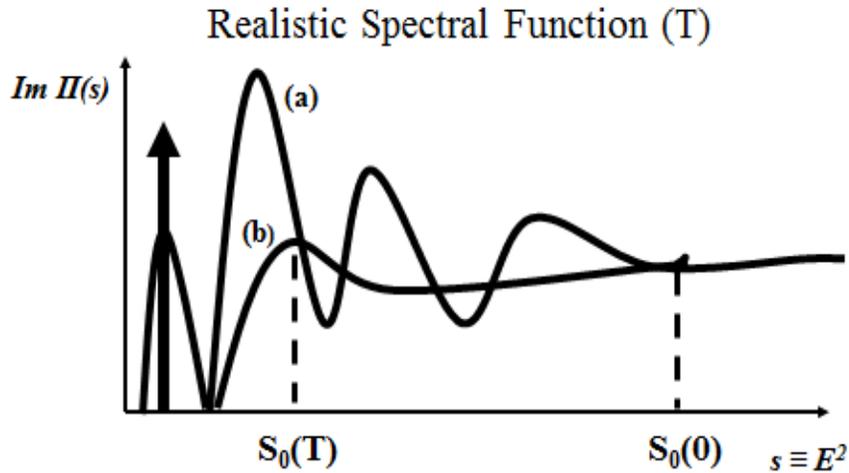}
		\caption{\small{Typical hadronic spectral function, $Im \Pi(s)$, at $T=0$, curve (a), showing a pole and three resonances. The squared energy $s_0(0)$ is the threshold for PQCD. At finite $T$, curve (b), the stable hadron develops a width, the resonances become broader (only one survives in this example), and the PQCD threshold, $s_0(T)$, approaches the origin. Eventually, at $ T\simeq T_c$, there will be no trace of resonances, and $s_0(T_c) \rightarrow 0$.}}
		\label{fig:figure1}
	\end{center}
\end{figure}
This could be e.g. the pion pole entering the axial-vector or the pseudo-scalar correlator, with the spectral function 

\begin{equation}
{\mbox{Im}}\, \Pi(s)|_{POLE} = 2 \, \pi f_\pi^2 \; \delta(s - M_\pi^2) \,,\label{Eq:ImPi}
\end{equation}

where $f_\pi \simeq 93\; {\mbox{MeV}}$ is the pion (weak-interaction) decay constant, defined as $\langle 0|A_\mu(0)| \pi(p)\rangle = \sqrt{2}\, f_\pi \, p_\mu$,
and $M_\pi$ its mass.
This is followed by resonances of widths increasing in size with increasing $s$, and corresponding to poles in the second Riemann sheet in the complex $s$-plane. For instance, for narrow resonances the Breit-Wigner parametrization is normally adequate

\begin{equation}
{\mbox{Im}}\, \Pi(s)|_{RES} = f_R^2 \,\, \frac{M_R^3 \; \Gamma_R}{(s - M_R^2)^2 + M_R^2 \Gamma_R^2} ;,\label{BW0}
\end{equation}

where $f_R$ is the coupling of the resonance to the current entering a correlation function, $M_R$ its mass, and $\Gamma_R$ its (hadronic) width. 
At high enough squared energy, $s_0(0) \simeq 2 -3 \; {\mbox{GeV}^2}$, the spectral function becomes smooth, and should be well approximated by perturbative QCD (PQCD). In the sequel, this parameter will be indistinctly referred to as the perturbative QCD threshold, the continuum threshold, or the deconfinement parameter.  At finite $T$ this spectrum gets distorted. The pole in the real axis moves down into the second Riemann sheet, thus generating a finite width. The widths of the rest of the resonances increase with increasing $T$, and some states begin to disappear from the spectrum, as first proposed in \cite{dimuonDL}. Eventually, close to, or at the critical temperature for de-confinement, $T \simeq T_c$, there will be no trace of the resonances, as their widths would be very large, and their couplings to hadronic currents would approach zero. At the same time, $s_0(T)$ would approach the origin. Thus $s_0(T)$ becomes a phenomenological order parameter for quark-de-confinement, as first proposed by Bochkarev and Shaposnikov \cite{BS}. This order parameter, associated with QCD de-confinement, is entirely phenomenological and quite different from the Polyakov-loop used by LQCD. Nevertheless, and quite importantly, qualitative and quantitative conclusions regarding this phase transition (or crossover), and the behaviour of QCD and hadronic parameters as $T \rightarrow T_c$ obtained from QCDSR and LQCD should agree. It is reassuring that this turns out to be the case, as will be reviewed here. In this scenario,
whatever happens to the mass is totally irrelevant; it could either increase or decrease with temperature, providing no information about de-confinement. The crucial parameters are the width and the coupling, but not the mass. In fact, if a particle mass would approach the origin, or even vanish with increasing temperature, this in itself is not sufficient to signal de-confinement, as a massless particle with a finite coupling and width would still contribute to the spectrum. What is required is that the widths diverge, and the couplings vanish. In all applications of QCD sum rules at finite $T$, the hadron masses in some channels decrease, and in other cases they  increase slightly with increasing $T$. At the same time, for all light, and heavy-light quark bound states the widths are found to diverge and the couplings to vanish close to or at $T_c$, thus signalling de-confinement. However, in the case of charmonium and bottonium hadronic states, after an initial surge, the widths decrease considerably with increasing temperature, while the couplings are initially independent of $T$, and eventually grow sharply close to $T_c$. This survival of charmonium states was first predicted from thermal QCD sum rules \cite{ccbar1}-\cite{ccbar2}, and later extended to bottonium \cite{bbar}, in qualitative agreement with LQCD \cite{LQCD_QQbar1}-\cite{LQCD_QQbar2}.\\
In addition to $s_0(T)$, there is another important thermal QCD quantity, the quark condensate, this time a fundamental order parameter of chiral-symmetry restoration. It is well known that QCD in the light-quark sector possesses a chiral $SU(2) \times SU(2)$ symmetry in the limit of zero-mass up- and down-quarks. This chiral-symmetry is realized in the Nambu-Goldstone fashion, as opposed to the Wigner-Weyl realization \cite{Pagels}. In other words, chiral-symmetry is a dynamical, as opposed to a classification symmetry. Hence, the pion mass squared vanishes as the quark mass

\begin{equation}
M_\pi^2 = B\; m_q \,, \label{Mpi2}
\end{equation}

and the pion decay constant squared vanishes as the quark-condensate

\begin{equation}
f_\pi^2 = \frac{1}{B}\; \langle \bar{q}\, q \rangle\,, \label{fpi2}
\end{equation}

where B is a constant. While from Eq.(\ref{Mpi2}) $M_\pi^2$ can vanish as the quark mass at $T=0$, the vanishing of $f_\pi$ can only take place at finite temperature, as $\langle \bar{q}\, q \rangle_T \rightarrow 0$ at $T=T_c$, the critical temperature for {\it{chiral-symmetry restoration}}. The correct meaning of {\it{chiral-symmetry restoration}} is a phase transition from a Nambu-Goldstone realization of chiral $SU(2) \times SU(2)$ to a Wigner-Weyl realization of the symmetry. In other words, there is a clear distinction between a symmetry of the Lagrangian and a symmetry of the vacuum \cite{Pagels}.\\
In summary, at finite temperature the  hadronic parameters to provide relevant information on de-confinement are the hadron width and its coupling to the corresponding interpolating current. On the QCD side we have (a)  the chiral condensate, $\langle \bar{q}\, q \rangle_T$, providing information on chiral-symmetry restoration, and (b) the onset of PQCD as determined by the squared energy, $s_0(T)$, providing information on quark-de-confinement. The next step is to relate these two sectors. This is currently done by considering the complex squared energy plane and invoking Cauchy's theorem, as described first at $T=0$ in Section II, and at finite $T$ in Section III. However, to keep a historical perspective, a summary of the original approach \cite{BS}, not entirely based on Cauchy's theorem, will be provided first.

\section{QCD sum rules at $T=0$}
The primary object in the QCD sum rule approach is the current-current correlation function

\begin{equation}
\Pi (q^{2})   = i \, \int\; d^{4} \, x \, e^{i q x} \,
\langle 0|T( J(x)   J^{\dagger}(0))|0\rangle \,,\label{correlator}
\end{equation}

where $J(x)$ is a local current built either from the QCD quark/gluon fields, or from hadronic fields. In the case of QCD, and invoking the Operator Product Expansion (OPE) of current correlators at short distances beyond perturbation theory \cite{SVZ}-\cite{Review1}, one of the two pillars of the QCD sum rule method, one has

\begin{equation}
\Pi(q^2)|_{\mbox{\scriptsize{QCD}}} = C_0 \, \hat{I} + \sum_{N=1} \frac{C_{2N} (q^2,\mu^2)}{(-q^2)^{N}} \langle \hat{\mathcal{O}}_{2N} (\mu^2) \rangle \;, \label{OPE}
\end{equation}

where $\langle \hat{\mathcal{O}}_{2N} (\mu^2) \rangle \equiv \langle0| \hat{\mathcal{O}}_{2N} (\mu^2)|0 \rangle$, $\mu^2$ is a renormalization scale, the Wilson coefficients $C_N$ depend on the Lorentz indexes and quantum numbers of the currents, and on the local gauge invariant operators ${\hat{\mathcal{O}}}_N$ built from the quark and gluon fields in the QCD Lagrangian. These operators are ordered by increasing dimensionality and the Wilson coefficients are calculable in PQCD. The unit operator above has dimension $d\equiv 2 N =0$ and $C_0 \hat{I}$ stands for the purely perturbative contribution. At $T=0$ the dimension $d\equiv 2 N = 2$ term in the OPE cannot be constructed from gauge invariant operators built from the quark and gluon fields of QCD (apart from negligible light-quark mass corrections). In addition, there is no evidence from such a term from analyses  using  experimental data \cite{D11}-\cite{D12}, so that the OPE starts at dimension $d \equiv 2 N = 4$.
\begin{figure}[ht]
\begin{center}
\includegraphics[height=3.0in, width=3.0in]{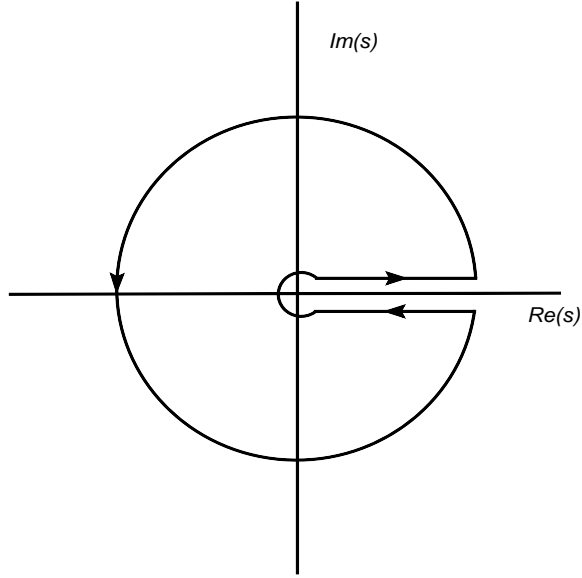}
  \caption{\small{The complex squared energy, $s$-plane, used in Cauchy's theorem. The discontinuity across the positive real axis is given by the hadronic spectral function, and QCD is valid on the circle of radius $s_0$, the threshold for PQCD.}}
\label{fig:figure2}
\end{center}
\end{figure}
The contributions at this dimension arise from the vacuum expectation values of the gluon field squared (gluon condensate), and of the quark-anti quark  fields (the quark condensate) times the quark mass.\\
While the Wilson coefficients in the OPE, Eq.(\ref{OPE}) can be computed in PQCD, the values of the vacuum condensates cannot be obtained analytically from first principles, as this would be tantamount to solving QCD analytically and exactly. These condensates can be determined from the QCDSR themselves, in terms of some input experimental information, e.g. spectral function data from $e^+ e^-$ annihilation into hadrons, or hadronic decays of the $\tau$-lepton. Alternatively, they may obtained by LQCD simulations. An exception is the value of the quark- condensate which is related to the pion decay constant through Eq.(\ref{fpi2}). 
As an example, let us consider the conserved vector current correlator

\begin{equation}
\Pi_{\mu\nu} (q^{2})   = i \, \int\; d^{4} \, x \, e^{i q x} \,
\langle 0|T( V_{\mu}(x)   V_{\nu}^{\dagger}(0))|0\rangle \,
= \,(-g_{\mu\nu}\, q^2 + q_\mu q_\nu) \, \Pi(q^2)  \; ,\label{VVcorrelator}
\end{equation} 

where $V_\mu(x) = \frac{1}{2}[: \bar{u}(x) \gamma_\mu \, u(x) - \bar{d}(x) \gamma_\mu \, d(x):]$ is the (electric charge neutral) conserved vector current in the chiral limit ($m_{u,d} = 0$), and $q_\mu = (\omega, \vec{q})$ is the four-momentum carried by the current. The function $\Pi(q^2)$ in PQCD is normalized as

\begin{equation}
	{\mbox{Im}} \,\Pi(q^2) = \frac{1}{8\pi}\left[ 1 + {\cal{O}} \left( \alpha_s(q^2)\right)\right] \,, \label{Pi}
\end{equation}

where the first term in brackets corresponds to the one-loop contribution, and  ${\cal{O}}(\alpha_s(q^2))$ stands for the multi-loop radiative corrections. The
leading non-perturbative term of dimension $d\equiv 2 N = 4$ is given by

\begin{equation}
C_4 \langle \hat{\mathcal{O}}_{4}  \rangle = 
\frac{\pi}{3} \langle \alpha_s G^2\rangle + 4 \pi^2 (\overline{m}_u+\overline{m}_d) \;\langle \bar{q} \,q\rangle\,,   \label{C4}
\end{equation}

a renormalization group invariant quantity, where $\overline{m}_{u,d}$ are the QCD current quark masses in the $\overline{MS}$ regularization scheme, and $\langle\overline{u} \,u\rangle = \langle\overline{d}\,d\rangle \equiv \langle\overline{q} \,q\rangle$. No radiative corrections to vacuum condensates will be considered here. The scale dependence of the quark condensate cancels with the corresponding dependence of the quark masses. In general, the numerical values of the vacuum condensates cannot be determined analytically from first principles, as mentioned earlier. An important exception is the quark condensate term above, whose value follows from the Gell-Mann-Oakes-Renner relation in chiral $SU(2) \times SU(2)$ symmetry \cite{GMOR}

\begin{equation}
(\overline{m}_u+\overline{m}_d) \;\langle \bar{q}\,q\rangle =  f_\pi^2 M_\pi^2 \;, \label{GMOR}
\end{equation}

where $f_\pi = 92.21 \pm 0.02 \, {\mbox{MeV}}$ is the experimentally measured pion decay constant \cite{PDG} . Corrections to this relation, essentially hadronic, are small and at the level of a few percent \cite{GMOR}.\\
Turning to the hadronic representation of the current correlation function $\Pi(q^2)$ in the time-like region, $q^2 \equiv s \geq 0$, in Eq.\eqref{VVcorrelator}, it is given by the rho-meson resonance at leading order. To a good approximation this is well described by a Breit-Wigner form

\begin{equation}
	\frac{1}{\pi} {\mbox{Im}} \Pi|_{HAD}(s) = \frac{1}{\pi} \; f_\rho^2 \, \frac{M_\rho^3 \,\Gamma_\rho}{\left(s - M_\rho^2\right)^2 + M_\rho^2 \,\Gamma_\rho^2}\;, \label{BW}
\end{equation}

where $f_\rho = 4.97 \,\pm\,0.07$ is the coupling of the $\rho$-meson to the vector current measured in its leptonic decay \cite{PDG}, and $M_\rho = 775.26 \,\pm\, 0.25\, {\mbox{MeV}}$, and $\Gamma_\rho = 147.8 \,\pm\, 0.9 \,{\mbox{MeV}}$ are the experimental mass and width of the $\rho$-meson, respectively. This 
parametrization has been normalized such that the area under it equals the area under a zero width expression, i.e. 

\begin{equation}
{\mbox{Im}} \,\Pi|^{(0)}_{HAD}(s) =  M_\rho^2 \; f_\rho^2 \; \delta(s - M_\rho^2) \,. \label{Delta}
\end{equation}

The next step is to find a way to relate the QCD representation of $\Pi(s)$ to its hadronic counterpart. Historically, at $T=0$, one of the first attempts was made in \cite{SVZ} using as a first step a dispersion relation (Hilbert transform), which follows from Cauchy's theorem in the complex squared energy $s$-plane 

\begin{equation}
\varphi_N(Q_0^2) \equiv \frac{1}{N !}\left(- \frac{d}{d Q^2}\right)^N \Pi(Q^2)|_{Q^2=Q_0^2} = \frac{1}{\pi}  \int_0^\infty \frac{Im\, \Pi(s)}{(s+Q_0^2)^{N+1}} \, ds \;, \label{Hilbert}
\end{equation}

where $N$ equals the number of derivatives required for the integral to converge asymptotically, $Q_0^2$ is a free parameter, and $Q^2 \equiv - q^2 > 0$. As it stands, the dispersion relation, Eq.(\ref{Hilbert}), is a tautology. In the early days of High Energy Physics the optical theorem was invoked in order to relate the spectral function, $Im \Pi(s)$, to a total hadronic cross section, together with some assumptions about its asymptotic behaviour, and thus relate the integral to the real part of the correlator, or its derivatives. The latter could, in turn, be related to e.g. scattering lengths. The procedure proposed in \cite{SVZ} was to parametrize the hadronic spectral function as

\begin{equation} 
Im \,\Pi(s)|_{HAD}= Im \,\Pi(s)|_{POLE} +\, Im \,\Pi(s)|_{RES} \,\theta(s_0-s) + \, Im\, \Pi(s)|_{PQCD}\,\theta(s-s_0) \;,\label{hadspec}
\end{equation}

where the ground state pole (if present) is followed by the resonances which merge smoothly into the hadronic continuum above some threshold $s_0$. This continuum is expected to be well represented by PQCD if $s_0$ is large enough. Subsequently, the left hand side of this dispersion relation is written in terms of the QCD OPE, Eq.(\ref{OPE}). The result is a sum rule relating hadronic to QCD information. Subsequently, in \cite{SVZ} a specific asymptotic limiting process in the parameters $N$ and $Q^2$ was performed, i.e. $\lim Q^2 \rightarrow \infty$ and $\lim N \rightarrow \infty$, with $Q^2/N \equiv M^2$ fixed, leading to Laplace transform QCD sum rules, expected to be more useful than the original Hilbert moments,  

\begin{equation}
\hat{L}_M [\Pi(Q^2)]\equiv\lim_{\stackrel{Q^2,N \rightarrow \infty}{Q^2/N\equiv M^2}} \frac{(-)^N}{(N-1)!} (Q^2)^N \left(\frac{d}{d Q^2} \right)^N \Pi(Q^2) 
\equiv \Pi(M^2) 
= \frac{1}{M^2} \int_0^{\infty} \frac{1}{\pi} Im \,\Pi(s) \, e^{-s/M^2} \, ds. \label{Laplace}
\end{equation}

Notice that this limiting procedure leads to the transmutation of $Q^2$ into the Laplace variable $M^2$.
This equation is still a tautology. In order to turn it into something with useful content one still needs to invoke Eq.(\ref{hadspec}). 
In applications of these sum rules \cite{Review1} $\Pi(M^2)$ was computed in QCD by applying the Laplace operator $\hat{L}_M$ to the OPE expression of $\Pi(Q^2)$, Eq.(\ref{OPE}),  and the spectral function on the right hand side was parametrized as in Eq.(\ref{hadspec}).  The function $\Pi(M^2)$ in PQCD involves the transcendental function  $\mu(t,\beta,\alpha)$ \cite{Erdelyi}, as first discussed in \cite{EdeR}. This feature, largely ignored for a long time, has no consequences in PQCD at the two-loop level. However, at higher orders, ignoring this relation leads to wrong results. It was only after the mid 1990's that this situation was acknowledged and higher order radiative corrections in Laplace transform QCDSR were properly evaluated.\\

This novel method had an enormous impact, as witnessed by the several thousand publications to date on analytic solutions to QCD in the non-perturbative domain \cite{Review1}. However, in the past decade, and as the subject moved towards high precision determinations to compete with  LQCD, these particular sum rules have fallen out of favour for a variety of reasons as detailed next. Last, but not least, Laplace transform QCDSR are ill-suited to deal with finite temperature, as explained below.

The first thing to notice in Eq.(\ref{Laplace}) is the introduction of an ad-hoc new parameter, $M^2$, the Laplace variable, which determines the squared energy regions where the exponential kernel would have a minor/major impact. It had been regularly advertised in the literature that a judicious choice of $M^2$ would lead to an exponential suppression of the often experimentally unknown resonance region beyond the ground state, as well as to a factorial suppression of higher order condensates in the OPE. In practice, though, this was hardly factually achieved, thus not supporting expectations. Indeed, since the parameter $M^2$ has no physical significance, other than being a mathematical artefact, results from these QCDSR would have to be independent of $M^2$ in a hopefully broad region. In applications, this so called stability window is often unacceptably narrow, and the expected exponential suppression of the unknown resonance region does not materialize. Furthermore, the factorial suppression of higher order condensates only starts at dimension $d=6$ with a mild suppression by a factor $1/\Gamma(3) = 1/2$. But beyond $d=6$ little, if anything, is numerically known about the vacuum condensates to profit from this feature. Another serious shortcoming of these QCDSR is that the role of the threshold for PQCD in the complex $s$-plane,  $s_0$, i.e. the radius of the circular contour in Fig. 2, is  exponentially suppressed. This is rather unfortunate, as $s_0$ is a parameter which, unlike $M^2$, has a clear physical interpretation, and which can be easily determined from data in some instances, e.g. $e^+ e^-$ annihilation into hadrons, $\tau$-lepton hadronic decays, etc.. When dealing with QCDSR at finite temperature, this exponential suppression of $s_0$ is utterly unacceptable, as $s_0(T)$ is {\bf{\it{the}}}  phenomenological order parameter of de-confinement!. A more detailed critical discussion of Laplace transform QCDSR may be found in \cite{QCDBook}. In any case, and due to the above considerations, no use will be made of these sum rules in the sequel.\\

A different attempt at relating QCD to hadronic physics was made by Shankar \cite{Shankar} (see also \cite{Mario1}-\cite{Mario3}) by considering the complex squared-energy s-plane  shown in Fig.2. The next step is the observation
that there are no singularities in this plane, except on the positive real axis where there might be a pole (stable particle) and a cut which introduces a discontinuity across this axis. This cut arises from the hadronic  resonances (on the second Riemann sheet) present in any given correlation function. Hence, from Cauchy's theorem in this plane (quark-hadron duality) one obtains 

\begin{equation}
\oint \Pi(s) \, ds = 0 = \int_{0}^{s_0}\Pi(s+i\epsilon)\, ds +\int_{s_0}^{0} \Pi(s-i \epsilon)\, ds 
+ \oint_{C(|s_0|)} \Pi(s)\, ds \,, \label{Cauchy}
\end{equation}

which becomes  finite energy sum rules (FESR)

\begin{equation}
 \int_{0}^{s_0} \frac{1}{\pi} \, {\mbox{Im}} \,\Pi(s)|_{HAD}\, P(s) \, ds =
 - \frac{1}{2 \pi i} \oint_{C(|s_0|)} \Pi(s)_{QCD}\, P(s) \,ds \,, \label{FESR}
\end{equation}

where an analytic function $P(s)$ has been inserted, without changing the result, and the radius of the circle $s=|s_0|$ is understood to be large enough for QCD to be valid there. The function $P(s)$ need not be an analytic function, in which case the contour integral instead of vanishing would be proportional to the residue(s) of the integrand at the pole(s). In some cases, this is deliberately considered, especially if the residue of the singularity is known independently, or conversely, if the purpose is to determine this residue. The  function $P(s)$ above is introduced in order to e.g. generate a set of FESR  projecting each and every vacuum condensate of different dimensionality in the OPE, Eq. \eqref{OPE}. For instance, choosing $P(s) = s^N$, with $N\geq 1$, leads to the FESR

\begin{equation}
(-)^{(N-1)} C_{2N} \langle {\mathcal{\hat{O}}}_{2N}\rangle = 8 \pi^2 \int_0^{s_0} ds\, s^{N-1} \,\frac{1}{\pi} \;{\mbox{Im}}\, \Pi(s)|_{\mbox{\scriptsize
{HAD}}}
- \frac{s_0^N}{N} \left[1+{\mathcal{O}}(\alpha_s)\right] \;\; (N=1,2,\cdots) \;,\label{FESR2}
\end{equation}

where the leading order vacuum condensates in the chiral limit ($m_q=0$) are the dimension $d \equiv 2 N = 4$ condensate, Eq.\eqref{C4},
and the dimension $d \equiv 2 N = 6$ four-quark condensate

\begin{equation}
C_6 \langle \hat{\mathcal{O}}_{6}  \rangle = - 8 \pi^3 \alpha_s \left[\langle (\bar{q} \gamma_\mu \gamma_5 \lambda^a q)^2 \rangle + \frac{2}{9} \langle(\bar{q} \gamma_\mu \lambda^a q)^2\rangle \right]
\,, \label{C6}
\end{equation}

where $\lambda_a$ are the SU(3) Gell-Mann matrices. A word of caution, first brought up in \cite{Shankar}, is important at this point, having to do with the validity of QCD on the circle of radius $|s_0|$ in Fig.2. Depending on the value of this radius, QCD may not be valid on the positive real axis, a circumstance called quark-hadron duality violation (DV). This is currently a contentious issue, which however has no real impact on finite temperature QCD sum rules, to wit. At $T=0$ one way to deal with potential DV is to introduce in the FESR, Eq.(\ref{FESR}), weight functions $P(s)$ which vanish on the positive real axis (pinched kernels) \cite{D11}-\cite{D12}, \cite{pinched11}-\cite{pinched12}, or alternatively, design specific models of duality violations \cite{pinched2}. The size of this effect is relatively small, becoming important only at higher orders (four- to five-loop order) in PQCD. Thermal QCD sum rules are currently studied only at leading one-loop order in PQCD, so that DV can be safely ignored. In addition, results at finite $T$ are traditionally  normalized to their $T=0$ values, so that only ratios are actually relevant.

In order to verify that the FESR, Eq.\eqref{FESR2}, give the right order of magnitude results, one can choose e.g. the vector channel,  use the zero-width approximation for the hadronic spectral function, ignore radiative corrections, and consider the $N=0$ FESR to determine $s_0$. The result is $s_0 \simeq 1.9 \, {\mbox{GeV}^2}$, or $\sqrt{s_0} \simeq 1.4 \,{\mbox{GeV}}$, which lies above the $\rho$-meson, and slightly below its very broad first radial excitation, $M_{\rho'} \simeq 1.5 \,{\mbox{GeV}}$. An accurate determination using the Breit-Wigner expression, Eq.\eqref{BW}, together with radiative corrections up to five-loop order in QCD, gives instead $s_0 = 1.44\, {\mbox{GeV}^2}$, or $\sqrt{s_0} = 1.2 \,{\mbox{GeV}}$, a very reassuring result. Among recent key applications of these QCD FESR are high precision determinations of the light- and heavy-quark masses \cite{QCDBook},\cite{quarkmassR1}-\cite{quarkmassR4}, now competing in accuracy with LQCD results, and the hadronic contribution to the muon magnetic anomaly $(g-2)_\mu$ \cite{g-21}-\cite{g-23}.\\

Turning to the case of heavy quarks, instead of FESR it is more convenient to use Hilbert moment sum rules \cite{RRY}, as described next. The starting point is the standard dispersion relation, or Hilbert transform, which follows from Cauchy's theorem in the complex $s$-plane, Eq.(\ref{Hilbert}).
In order to obtain practical information one invokes Cauchy's theorem in the complex $s$-plane (quark-hadron duality), so that the Hilbert moments, Eq.(\ref{Hilbert}) become effectively FESR

\begin{equation}
\varphi_N(Q_0^2)|_{HAD} =  \varphi_N(Q_0^2)|_{QCD} \;,\label{duality}
\end{equation}
where

\begin{equation}
\varphi_N(Q_0^2)|_{HAD} \equiv \frac{1}{\pi}\, 
\int_{0}^{s_0} \; \frac{ds}{(s+Q_0^2)^{(N+1)}} \,  Im \,\Pi(s)|_{HAD}\; , 
\end{equation}

\begin{equation}
 \varphi_N(Q_0^2)|_{QCD} \equiv \frac{1}{\pi}\, 
 \int_{4m_Q^2}^{s_0} \, \frac{ds}{(s+Q_0^2)^{(N+1)}} \,  Im \,\Pi(s)|_{PQCD}\; + \varphi_N(Q_0^2)|_{NP}\,.\label{HilbertQCD}
\end{equation}

In principle these sum rules are not valid for all values of the free parameter $Q_0^2$. In practice, though, a reasonably wide and stable window is found allowing for predictions to be made \cite{RRY}. Traditionally, these sum rules have been used in applications involving heavy quarks (charm, bottom), while FESR are usually restricted to the light-quark sector. However, there is no a-priori reason against departing from this approach. In the light-quark sector the large parameter is $Q^2$ (and $s_0$, the onset of PQCD), with the quark masses being small at this scale. Hence the PQCD expansion involves naturally inverse powers of $Q^2$. In the heavy-quark sector there is knowledge of PQCD in terms of the expansion parameter $Q^2/m_q^2$, leading to power series expansions in terms of this ratio. Due to this, most applications of QCDSR have been restricted to FESR in the light-quark sector, and Hilbert transforms for heavy quarks.\\

The non-perturbative moments above, $\varphi_N(Q_0^2)|_{NP} $, involve the vacuum condensates in the OPE, Eq.(\ref{OPE}). One important difference is that there is no quark condensate, as there is no underlying chiral-symmetry for heavy quarks. The would-be quark condensate $\langle \bar{Q}\, Q\rangle$ reduces to the gluon condensate, e.g. at leading order in the heavy-quark mass, $m_Q$ one has \cite{SVZ}

\begin{equation}
\big\langle \bar{Q} Q \big\rangle
= - \frac{1}{12\,m_Q}  \, \bigg\langle \frac{\alpha_s}{\pi}  \; G^2 \bigg\rangle\,,\label{QbarQ}
\end{equation}

where $m_Q$ is the heavy-quark mass (charm, bottom). Writing several FESR one obtains e.g. information on heavy-quark hadron masses, couplings and hadronic widths. Alternatively, using some known hadronic information one can find the values of QCD parameters, such as e.g. heavy-quark masses \cite{QCDBook},\cite{quarkmassR1}-\cite{quarkmassR4}, and the gluon condensate \cite{GGc1}-\cite{GGc2}. For a review see e.g. \cite{RRY}. Their extension to finite temperature will be discussed in Section 7.\\
The techniques required to obtain the QCD expressions of current correlators, both perturbative as well as non-perturbative (vacuum condensates) at $T=0$ are well described in detail in \cite{PT}.

\section{Light-quark axial-vector current correlator at finite $T$: relating de-confinement with chiral-symmetry restoration}

The first thermal QCDSR analysis was performed by Bochkarev and  Shaposhnikov  in 1986 \cite{BS}, using mostly the light-quark vector current correlator ($\rho$- and $\phi$-meson channels) at finite temperature, in the framework of Laplace-transform QCD sum rules. 
Additional field-theory support for such an extension was given later in \cite{QFTT}, in response to baseless criticisms of the method at the time.
Laplace transform  QCDSR were in fashion in those days \cite{Review1}, but their extension to finite $T$ turned out to  be a major breakthrough, opening up a new area of research (for early work see e.g. \cite{VARIOUS1}-\cite{VARIOUS8}). The key results of this pioneer paper \cite{BS} were the temperature dependence of the masses of the $\rho$ and the $\phi$ vector mesons, as well as the threshold for PQCD, $s_0(T)$. With hindsight, instead of the vector mesons masses, it would have been better to determine the vector meson couplings to the vector current. EHowever, at the time there were some proposals to consider the hadron masses as relevant thermal parameters. We have known for a long time now, that this was an ill-conceived idea. In fact, the $T$-dependence of hadron masses is irrelevant for the description of the behaviour of QCD and hadronic matter, and the approach to de-confinement and chiral-symmetry restoration. This was discussed briefly already in Section 1, and in more detail below. Returning to \cite{BS}, their results for the $T$-dependence of $s_0(T)$, i.e. the de-confinement phenomenological order parameter, clearly showed a sharp decrease with increasing $T$. Indeed, $s_0(T)$ dropped from $s_0(0) \simeq 2 \, {\mbox{GeV}^2}$, to $s_0(T_c) \simeq 0.2 \, {\mbox{GeV}^2}$ at $T_c \simeq 150 \,{\mbox{MeV}}$. A similar behaviour was also found in the $\phi$-meson channel. The masses in both cases had decreased only by some $10 \%$.\\
\begin{figure}[ht!]
	\begin{center}
		\includegraphics[height=3.2in, width=4.7in]{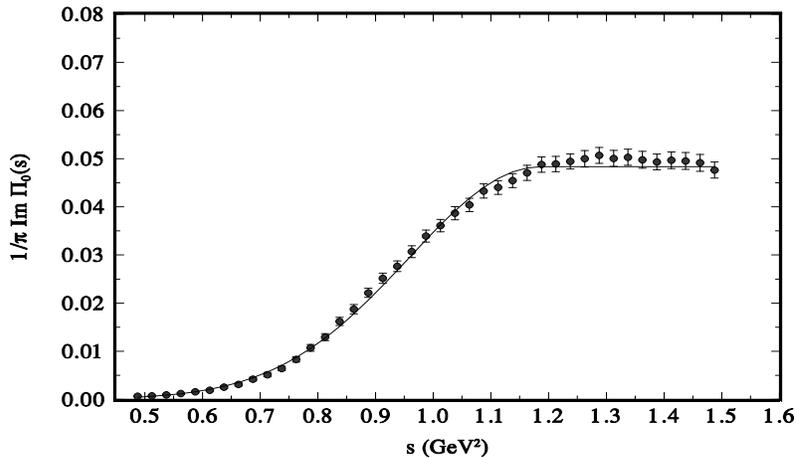}
		\caption{\small{The experimental data points of the axial-vector spectral function from the ALEPH Collaboration \cite{ALEPH3}, together with the fit using Eq.(\ref{A1}) (solid curve).}}.
		\label{fig:figure3}
	\end{center}
\end{figure}

The first improvement of this approach was proposed in \cite{FESRT1}, where QCD-FESR, instead of Laplace transform QCDSR, were used for the first time. The choice was the light-quark axial-vector correlator 

\begin{equation}
\Pi_{\mu\nu} (q^{2})   = i \, \int\; d^{4} \, x \, e^{i q x} \,
\langle 0|T( A_{\mu}(x)   A_{\nu}^{\dagger}(0))|0 \rangle \,
= \,-g_{\mu\nu}\, \Pi_1(q^2)  \;  + q_\mu q_\nu \, \Pi_0(q^2)  \; ,\label{AAcorrelator}
\end{equation} 

where $A_\mu(x) = : \bar{u}(x) \gamma_\mu \, \gamma_5 \, d(x):$ is the (electrically charged) axial-vector current,  and $q_\mu = (\omega, \vec{q})$ is the four-momentum carried by the current. The functions $\Pi_{0.1}(q^2)$  are free of kinematical singularities, a key property needed in writing dispersion relations and sum rules, with $\Pi_0(q^2)$ normalized as

\begin{equation}
{\mbox{Im}} \,\Pi_0(q^2)|_{QCD} = \frac{1}{4\pi}\left[ 1 + {\cal{O}} \left( \alpha_s(q^2)\right)\right] \,. \label{Pi0Axial}
\end{equation}
\begin{figure}[ht!]
	\begin{center}
		\includegraphics[height=3.2in, width=4.7in]{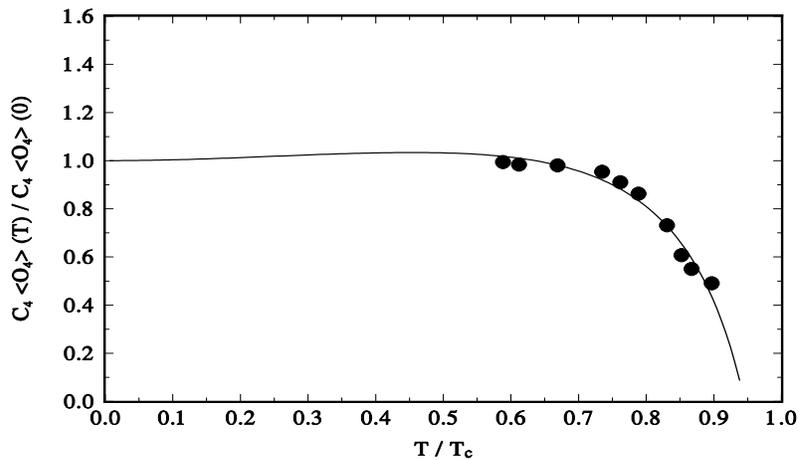}
		\caption{\small{The normalized thermal behaviour of the gluon condensate (solid curve), together  with LQCD results  (dots) \cite{LQCDGG} for $T_c = 197 \, {\mbox{MeV}}$.}}.
		\label{fig:figure4}
	\end{center}
\end{figure}

Notice the difference in a factor-two with the  normalization in Eq.(\ref{Pi}). This is due to the currents in Eq.(\ref{AAcorrelator}) being electrically charged, and those in Eq.(\ref{VVcorrelator}) being electrically neutral (thus involving an overall factor $1/2$, as stated after Eq.(\ref{VVcorrelator}))
The reason for this choice of correlation function was that since  the axial-vector correlator involves the pion decay constant, $f_\pi$ on the hadronic sector, the thermal FESR would provide a relation between $f_\pi(T)$ and $s_0(T)$. Since the former is related to the quark condensate $\langle \bar{q} q \rangle(T)$, Eq.(\ref{fpi2}), one would then obtain a relation between chiral-symmetry restoration and de-confinement, the latter being encapsulated in $s_0(T)$. A very recent
study \cite{Polyakov} of the relation between $s_0(T)$ and the trace of the Polyakov loop, in the framework of a non-local $SU(2)$ chiral quark model, concludes that both parameters provide the same information on the deconfinement phase transition. This conclusion holds for both zero and finite chemical potential. This result validates the thirty-year old phenomenological assumption of \cite{BS}, and its subsequent use in countless thermal QCD sum rule applications.
We will first assume  pion-saturation of the hadronic spectral function, in order to follow closely \cite{FESRT1}. Subsequently, we shall describe recent precision results in this channel \cite{FESRTAA}. Starting at $T=0$, the pion-pole contribution to the hadronic spectral function in the FESR, Eq.(\ref{FESR2}), is given by

\begin{equation}
{\mbox{Im}} \,\Pi_0(q^2)|_{HAD} = 2 \, \pi \, f_\pi^2 \, \delta(s) \,,\label{PionPole0}
\end{equation}

where $\delta(s - m_\pi^2)$ above was approximated in the chiral limit. With $C_2 \langle {\mathcal{\hat{O}}}_{2}\rangle = 0$ (see Eq.(\ref{OPE})),  the first FESR, Eq.(\ref{FESR2}) for $N=1$ simply reads

\begin{equation}
s_0 \, = \, 8 \, \pi^2 \, f_\pi^2 \,. \label{FESRT0}
\end{equation}

\begin{figure}[ht!]
	\begin{center}
		\includegraphics[height=3.2in, width=4.7in]{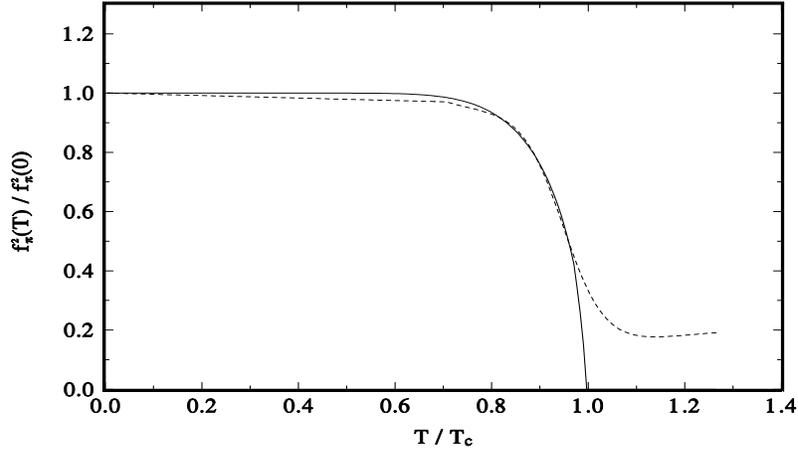}
		\caption{\small{The quark condensate $\langle \bar{q} q \rangle (T)/\langle \bar{q} q \rangle(0) = f_\pi^2(T)/f_\pi^2(0)$ as a function of $T/T_c$ in the chiral limit ($m_q=M_\pi=0$) with $T_c = 197\; {\mbox{MeV}}$ \cite{LQCDqq1} (solid curve), and for finite quark masses from a fit to lattice QCD results \cite{LQCDqq2}-\cite{LQCDqq3} (dotted curve)}}.
		\label{fig:figure5}
	\end{center}
\end{figure}

Numerically, $s_0 \simeq 0.7 \, {\mbox{GeV}^2}$, which is a rather small value, the culprit being the pion-pole approximation to the spectral function. In fact, as it will be clear later, when additional information is incorporated into Eq.(\ref{PionPole0}) in the form of the next hadronic state, the $a_1(1260)$, the value of $s_0$ increases substantially. In any case, thermal results will be normalized to the $T=0$ values.\\
The next step in \cite{FESRT1} was to use the Dolan-Jackiw \cite{DJ} thermal quark propagators, equivalent to the Matsubara formalism at the one-loop level, to find the QCD and hadronic spectral functions. For instance, at the QCD one-loop level the thermal quark propagator becomes

\begin{equation}
S_F(k,T) = \frac{i}{\slashed{k} - m} \, - \,  \frac{2 \, \pi}{\left(e^{|k^0|/T} + 1\right)} \; (\slashed{k}+ m) \; \delta(k^2 - m^2) \;, \label{DolanJackiw}
\end{equation}
 
and an equivalent expression for bosons, except for a positive relative sign between the two terms above, and the obvious replacement of the Fermi by the Bose thermal factor. An advantage of this expression is that it allows for a straightforward calculation of the imaginary part of current correlators, which is the function entering QCDSR.  It turns out that there are two distinct thermal contributions, as first pointed out in \cite{BS}. One in the time-like region $s = q^2 \geq 0$, called the annihilation term, and another in the space-like region  $s = q^2 \leq 0$, referred to as the scattering term. Here, $q^2 = \omega^2 - |\bf{q}^2|$, where $\omega$ is the energy, and $\bf{q}$ the three-momentum with respect to the thermal bath. The scattering term can be visualized as due to the scattering of quarks and hadrons, entering spectral functions, with quarks and hadrons in the hot thermal bath. In the complex energy $\omega$-plane, the correlation functions have cuts in both the positive as well as the negative real axes, folding into one single cut along the positive real axis in the complex $s = q^2$ plane. These singularities survive at $T=0$. On the other hand, the space-like contributions, non-existent at $T=0$,  if present at $T \neq 0$ are  due to cuts in the $\omega$-plane, centred at $\omega = 0$, with extension $-|\bf{q}| \leq \, \omega \, \leq|\bf{q}|$. In the limit $|\bf{q}| \rightarrow 0$, i.e. in the rest-frame of the medium, this contribution either vanishes entirely, or it becomes proportional to a delta-function $\delta(\omega^2)$ in the spectral function, depending on the $q^2$ behaviour of the current correlator. A detailed derivation of a typical scattering term is done in the Appendix.\\
Proceeding to finite $T$, the thermal version of the  QCD spectral function, Eq.(\ref{Pi0Axial}), in the time-like (annihilation) region, and in the chiral limit ($m_q=0$) becomes

\begin{equation}
{\mbox{Im}} \,\Pi^{a}_0(\omega,T)|_{QCD} = \frac{1}{4 \pi} \, \left[ 1 - 2 \, n_F \left(\frac{\omega}{2\,T} \right) \right] \,\, \theta(\omega^2) = \frac{1}{4 \pi} \: \tanh\left(\frac{\omega}{4 \,T}\right) \, \theta(\omega^2) \,, \label{Pi0A+T}
\end{equation} 

and the counterpart in the space-like (scattering) region

\begin{equation}
{\mbox{Im}} \,\Pi^{s}_0(\omega,T)|_{QCD} = \frac{4}{\pi} \, \delta(\omega^2) \int_0^{\infty} y \, n_F\left( \frac{y}{T}\right) \,dy \, = \frac{\pi}{3} \, T^2 \, \delta (\omega^2) \, \label{Pi0A-T}
\end{equation} 

where $n_F(z) = 1/(1 + e^z)$ is the Fermi thermal factor. A detailed derivation for finite quark masses is given in the Appendix. On the hadronic side, the scattering term at leading order is a two-loop effect, as the axial-vector current couples to three pions. This contribution is highly phase-space suppressed, and can be safely ignored.  The leading order thermal FESR is then

\begin{equation}
8\, \pi^2 \, f_\pi^2 (T) = \frac{4}{3} \, \pi^2 \, T^2 + \, \int_0^{s_0(T)} ds \left[ 1 - 2 \, n_F \left(\frac{\sqrt{s}}{2T} \right) \right] \,\label{FESRT}
\end{equation}

which relates chiral-symmetry restoration, encapsulated in $f_\pi^2(T) \propto - \langle \bar{q} \, q \rangle(T)$, with de-confinement as described by $s_0(T)$. At the time of this proposal \cite{FESRT1} there was no LQCD information on the thermal behaviour of the quark condensate (or $f_\pi$). One source of information on $f_\pi(T)$  was available from chiral perturbation theory, CHPT, \cite{fpiTCHPT}, whose proponents claimed it was valid up to intermediate temperatures. Using this information the de-confinement parameter  $s_0(T)$ was thus obtained in \cite{FESRT1}. It  showed  a monotonically decreasing behaviour with temperature, similar to that of $f_\pi(T)$, but vanishing at a much lower temperature. Quantitatively, this was somewhat disappointing, as it was expected  both critical temperatures to be similar. The culprit turned out to be the CHPT temperature behaviour of $f_\pi(T)$, which contrary to those early claims, is now known to be valid only extremely close to $T=0$, say only a few $\mbox{MeV}$. Shortly after this proposal \cite{FESRT1} the thermal behaviour of $f_\pi(T)$, valid in the full  temperature range, as obtained in \cite{Gatto1}-\cite{Gatto3},  was used in \cite{Gatto4} to solve the FESR, Eq.(\ref{FESRT}). The result showed a remarkable agreement between the ratio $f_\pi(T)/f_\pi(0)$ and $[s_0(T)/s_0(0)]$ over the whole range $T =( 0 - T_c)$. This result was very valuable, as it supported the method. Formal theoretical validation has been obtained recently in \cite{Polyakov}.

\begin{figure}[hb!]
	\begin{center}
	\includegraphics[height=3.2in, width=4.7in]{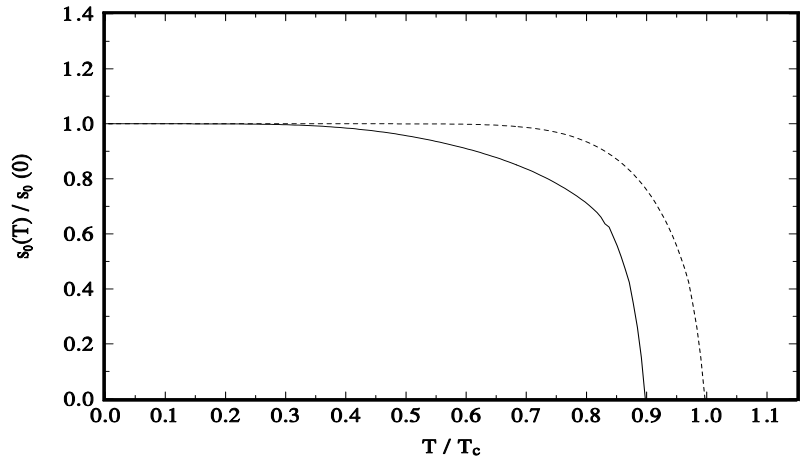}\caption{{\protect\small
			{Results from the FESR, Eqs.(\ref{FESRA1})-(\ref{FESRA3}), for the continuum threshold $s_0(T)/s_0(0)$ in the light-quark axial-vector channel, signalling de-confinement, (solid curve),  as a function of $T/T_c$, together with $f^2_\pi(T)/f^2_\pi(0) = \langle\bar{q} q\rangle(T)/\langle\bar{q} q\rangle(0)$ signalling chiral-symmetry restoration (dotted curve).}}}
	\label{figure6}
	\end{center}
\end{figure}

Further improved results along these lines were obtained more recently \cite{FESRTAA}, as summarized next.\\
\begin{figure}[ht!]
	\begin{center}
	\includegraphics[height=3.2in, width=4.7in]{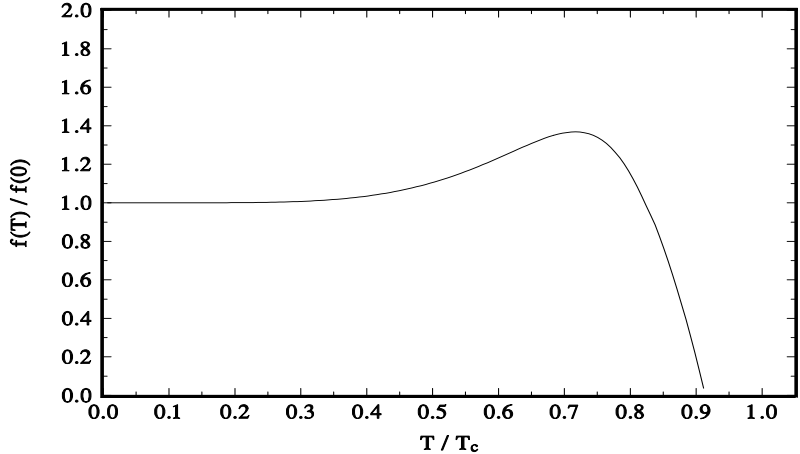}\caption{{\protect\small
			{Results from the FESR, Eqs.(\ref{FESRA1})-(\ref{FESRA3}), for the coupling of the $a_1(1260)$ resonance $f(T)/f(0)$  as a function of $T/T_c$.}}}
	\label{figure7}
		\end{center}
\end{figure}
The first improvement on the above analysis is the incorporation into the hadronic spectral function of the axial-vector three-pion resonance state, $a_1(1260)$. At $T=0$ there is ample experimental information in this kinematical region from hadronic decays of the $\tau$-lepton, as measured by the ALEPH Collaboration \cite{ALEPH1}-\cite{ALEPH3}. Clearly, there is no such information at finite $T$. The procedure is to first fit the data on the spectral function using some analytical expression involving hadronic parameters, e.g. mass, width, and coupling to the axial-vector current entering the current correlator. Subsequently, the QCDSR will fix the temperature dependence of these parameters together with that of $s_0(T)$. 
An excellent fit to the data (see Figure 3) was obtained in  \cite{FESRTAA} with the function

\begin{equation}
\frac{1}{\pi} \,{\mbox{Im}} \Pi_0(s)|_{HAD}= 2 \, f_\pi^2 \, \delta(s) \, + \, Cf\,
 \exp \left[- \left(\frac{s - M_{a_1}^2}{\Gamma^2_{a_1}} \right)^2  \right] \,, \label{A1}
 \end{equation}
 
where $M_{a_1} = 1.0891 {\mbox{GeV}}$, $\Gamma_{a_1} = 568.78\, {\mbox{MeV}}$, are the experimental values \cite{PDG},  and $Cf= 0.048326$ a fitted parameter. Notice that there is a misprint of Eq.(\ref{A1}) in \cite{FESRTAA}, where the argument of the exponential was not squared. Calculations there were done with the correct expression, Eq.(\ref{A1}). The dimension $d\equiv 2 N = 4$
condensate entering the FESR is given in Eq.(\ref{C4}), after multiplying by a factor-two to account for the different correlator normalization. The next term in the OPE, Eq.(\ref{OPE}), is the dimension $d\equiv 2 N = 6$ condensate, Eq.(\ref{C6}). As it stands, it is useless, as it cannot be determined theoretically. It has been traditional to invoke the so-called vacuum saturation approximation \cite{SVZ}, a procedure to saturate the sum over intermediate states by the vacuum state, leading to

\begin{equation}
C_6 \langle \hat{\mathcal{O}}_{6}  \rangle|_{A} \, \propto \, \alpha_s |\langle \bar{q}\, q \rangle|^2\,, \label{C6VVS}
\end{equation}

which is channel dependent, and has a very mild dependence on the renormalization scale. The numerical coefficient above is not important, as it cancels out in the ratio with respect to $T=0$. This  approximation has no solid theoretical justification, other than its simplicity. Hence, there is no reliable way of estimating corrections, which in fact appear to be rather large from comparisons between Eq. (\ref{C6VVS}) and direct determinations from data \cite{DS1}-\cite{DS2}. This poses no problem for the finite temperature analysis, where Eq.(\ref{C6VVS}) is only used to normalize results at $T=0$, and one is usually interested in the behaviour of ratios.
Next, the pion decay constant, $f_\pi$ is related to the quark-condensate through the Gell-Mann-Oakes-Renner relation
\begin{equation}
2\,f_\pi^2\,m_\pi^2 = - (m_u + m_d)\langle 0| \bar{u} u + \bar{d} d|0\rangle\;. \label{GMOR}
\end{equation} 

Chiral corrections to this relation are at the 5\% level \cite{GMOR}, and at finite $T$ deviations are negligible except very close to the critical temperature \cite{GMORT}.\\

\begin{figure}[ht!]
	\begin{center}
	\includegraphics[height=3.2in, width=4.7in]{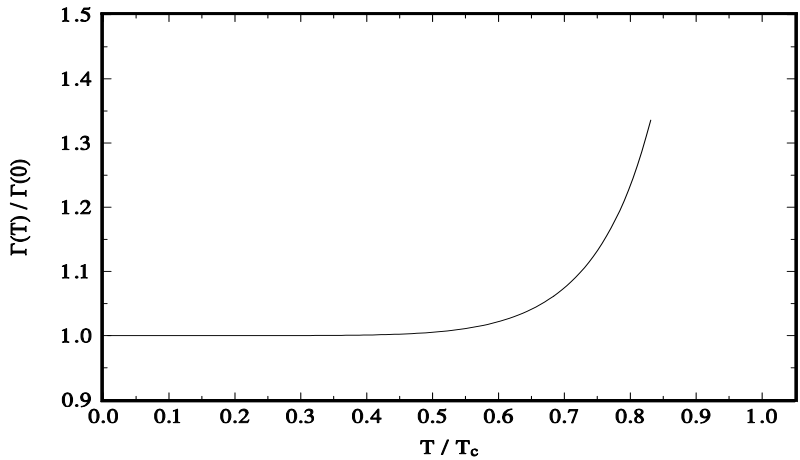}\caption{{\protect\small
			{Results from the FESR, Eqs.(\ref{FESRA1})-(\ref{FESRA3}), for the hadronic width of the $a_1(1260)$ resonance $\Gamma_{a_1}(T)/\Gamma_{a_1}(0)$  as a function of $T/T_c$.}}}
	\label{figure8}
		\end{center}
\end{figure}

Starting at $T=0$, the first three FESR, Eq.(\ref{FESR2}),  after dividing by a factor-two the first term on the right-hand-side, can be used to determine $s_0(0)$, and the $d\equiv 2 N = 4,6$ condensates. These values will be used later to normalize all results at finite $T$. The value thus obtained for $s_0(0)$ is $s_0(0) = 1.15 \,{\mbox{GeV}^2}$, a far more realistic result than that from using only the pion-pole contribution, Eq.(\ref{FESRT0}). Next, the $T=0$ values of the $d\equiv 2 N = 4,6$ condensates obtained from the next two FESR are in good agreement with determinations from data \cite{DS1}-\cite{DS2}.\\ Moving to finite $T$, in principle there are six unknown quantities to be determined from three FESR, to wit, (1) $s_0(T)$, (2) $f(T)$, and (3) $\Gamma_{a_1}(T)$ on the hadronic side, and (4) $f_\pi(T) \propto - \langle \bar{q} \, q \rangle (T)$, (5) $C_4 \langle \hat{\mathcal{O}}_{4}  \rangle =
\frac{\pi}{3} \langle \alpha_s G^2\rangle$ (in the chiral limit), and (6) $C_6 \langle \hat{\mathcal{O}}_{6}  \rangle$ on the QCD side. The latter can be determined using vacuum saturation, thus leaving five unknown quantities, for which there are three FESR. In \cite{FESRTAA} the strategy was to use LQCD results for the thermal quark  and gluon condensates, thus allowing the determination of $s_0(T)$, $f(T)$, and $\Gamma_{a_1}(T)$ from the three FESR.
The LQCD results are shown in Fig.4 for the gluon condensate \cite{LQCDGG}, and in Fig.5 for the quark condensate \cite{LQCDqq1}-\cite{LQCDqq3}.

The three FESR to be solved are then

\begin{equation}
	8  \pi^2 f^2_\pi(T) = \frac{4}{3}  \pi^2  T^2  + \int_0^{s_0(T)}ds \,\left[1 - 2\, n_F \left(\frac{\sqrt{s}}{2 T} \right) \right] 
	- 4 \,\pi^2\, \int_0^{s_0(T)} ds\,  \frac{1}{\pi}\, {\mbox{Im}}\, \Pi_0(s,T)|_{a_1}
	\;, \label{FESRA1}
\end{equation}

\begin{equation}
	- C_{4}\langle {\mathcal{\hat{O}}}_{4}\rangle(T) = 4 \pi^2 \int_0^{s_0(T)} ds\, s \frac{1}{\pi} {\mbox{Im}}\, \Pi_0(s)|_{a_1}
	-  \int_0^{s_0(T)}ds \, s \left[1 - 2  n_F\left(\frac{\sqrt{s}}{2 T}\right)\right] ,\label{FESRA2}
\end{equation}

\begin{equation}
	 C_{6}\langle {\mathcal{\hat{O}}}_{6}\rangle(T) = 4 \pi^2 \int_0^{s_0(T)} ds\, s^2 \frac{1}{\pi} {\mbox{Im}}\, \Pi_0(s)|_{a_1}
	-  \int_0^{s_0(T)}ds \; s^2 \left[1 - 2  n_F\left(\frac{\sqrt{s}}{2 T}\right)\right] \;.\label{FESRA3}
\end{equation}

The result for $s_0(T)$ is shown in Fig.6, together with that of $f_\pi(T)$, both normalized to their values at $T=0$. The difference between the behaviour of the two quantities lies well within the accuracy of the method. In fact, the critical temperatures for chiral-symmetry restoration and for de-confinement differ by some 10\%. In any case, it is reassuring that de-confinement precedes chiral-symmetry restoration, as expected from general arguments \cite{BS}. Next, the behaviour of
the $a_1(1260)$ resonance coupling to the axial-vector current, $f(T)$, is shown in Fig.7. As expected, it vanishes sharply as $T \rightarrow T_c$. The $a_1(1260)$ resonance width is shown in Fig.8. One should recall that at $T=0$ this resonance is quite broad, effectively some $500 \;{\mbox{MeV}}$, as seen from Fig.3. Hence, a 30\% increase in width, as indicated in Fig.8, together with the vanishing of its coupling, renders this resonance unobservable.\\
This completes the thermal analysis of the light-quark axial-vector channel, and we proceed to study the thermal behaviour of the corresponding vector channel.

\begin{figure}[ht!]
	\begin{center}
	\includegraphics[height=3.2in, width=4.7in]{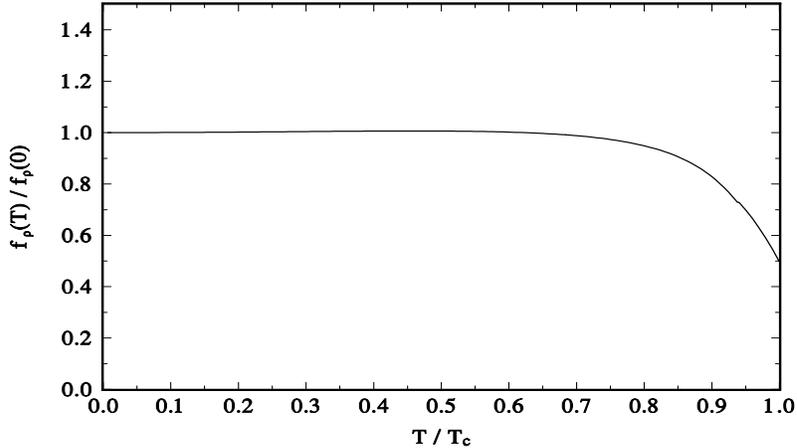}\caption{{\protect\small
			{Results from the FESR in the vector channel for the hadronic coupling  of the $\rho$-meson, $f_\rho(T)/f_\rho(0)$,  Eq.(\ref{frhoT}), as a function of $T/T_c$.}}}
	\label{figure9}
		\end{center}
\end{figure}
\begin{figure}[ht!]
	\begin{center}
	\includegraphics[height=3.2in, width=4.7in]{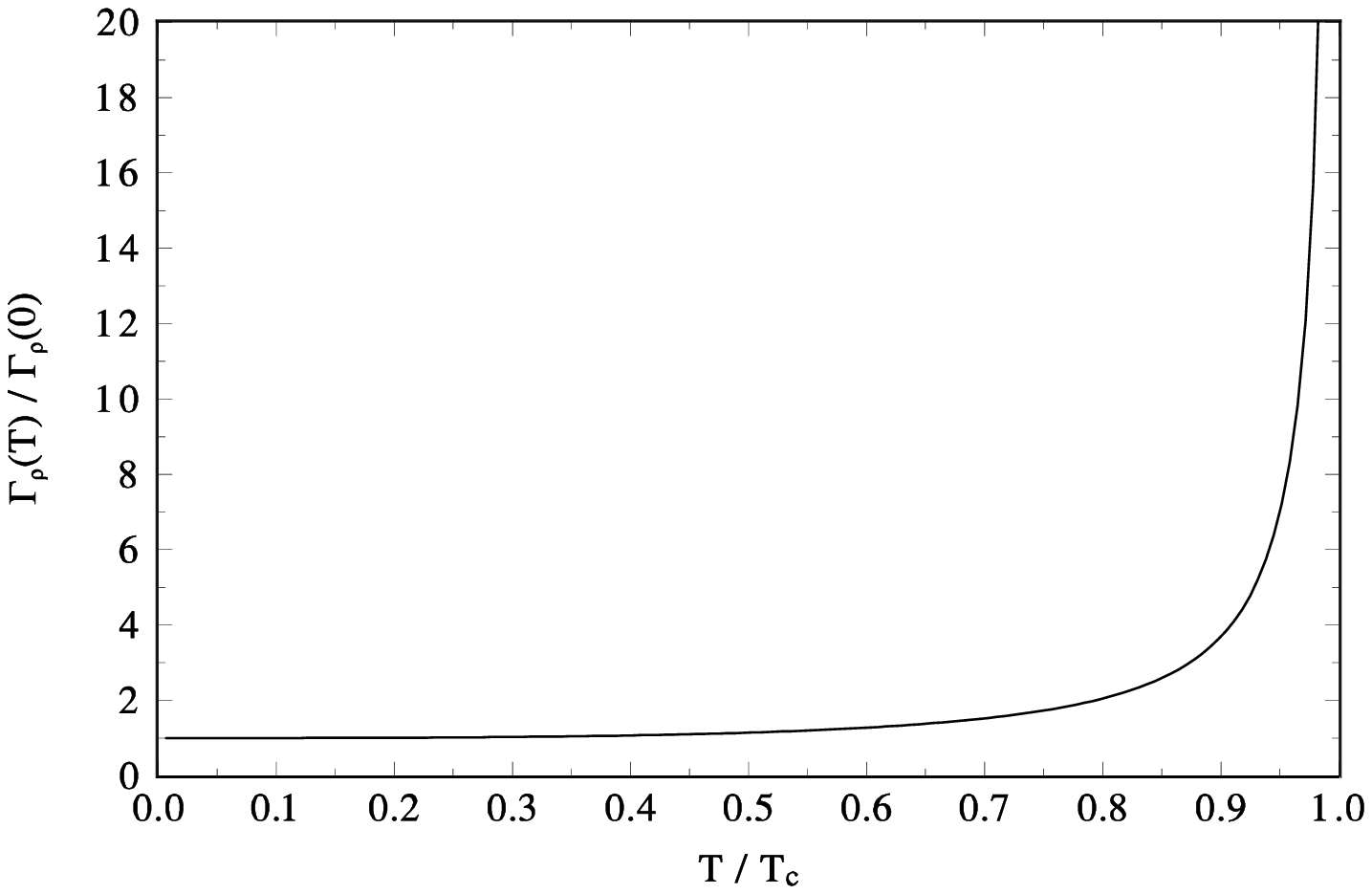}\caption{{\protect\small
			{Results from the FESR in the vector channel for the hadronic width  of the $\rho$-meson, $\Gamma_\rho(T)/\Gamma_\rho(0)$, Eq.(\ref{GAMMAf}), as a function of $T/T_c$.}}}
	\label{figure10}
		\end{center}
\end{figure}

\section{Light-quark vector current correlator at finite temperature and dimuon production in heavy-ion collisions at high energy}

The finite $T$ analysis in the vector channel \cite{rhoT} follows closely that in the axial-vector channel, except for the absence of the pion pole. However, we will summarize the results here as they have an important impact on the dimuon production rate in heavy nuclei collisions at high energies, to be discussed subsequently. This rate can be fully predicted using the QCDSR results for the $T$-dependence of the parameters entering the vector channel, followed by an extension to finite chemical potential (density).\\
Beginning with the QCD sector, the annihilation and scattering
spectral functions, in the chiral limit, are identical to those in the axial-vector channel, Eqs.(\ref{Pi0A+T})-(\ref{Pi0A-T}). An important difference is due to the presence of a hadronic scattering term, a  leading order two-pion one-loop, instead of a three-pion two-loop as in the axial-vector channel. This is given by \cite{rhoT}

\begin{equation}
\frac{1}{\pi} \,{\mbox{Im}}\, \Pi^{s}|_{HAD}(\omega,T) = \frac{2}{3\pi^2} \; \delta(\omega^2) \;\int_0^\infty y\, n_B\left(\frac{y}{T}\right) \,dy \;,\label{Pi-V}
\end{equation}

where $n_B(z) = 1/(e^{z}-1)$ is the Bose thermal function. Once again, there are three FESR, Eq.(\ref{FESR2}), to determine six quantities, $f_\rho(T)$, $M_\rho(T)$, $\Gamma_\rho(T)$, $s_0(T)$, $C_{4}\langle {\mathcal{\hat{O}}}_{4}\rangle(T)$, and $C_{6}\langle {\mathcal{\hat{O}}}_{6}\rangle(T)$. Starting with the latter, it can be related to the quark condensate in the vacuum saturation approximation \cite{SVZ}

\begin{equation}
C_6 \langle \hat{\mathcal{O}}_{6}  \rangle|_V \, \propto \,-\, \alpha_s\, |\langle \bar{q}\, q \rangle|^2\,, \label{C6AVS}
\end{equation}
where the sign is opposite to that in the axial-vector channel, Eq.(\ref{C6VVS}).

\begin{figure}[ht!]
	\begin{center}
	\includegraphics[height=3.2in, width=4.7in]{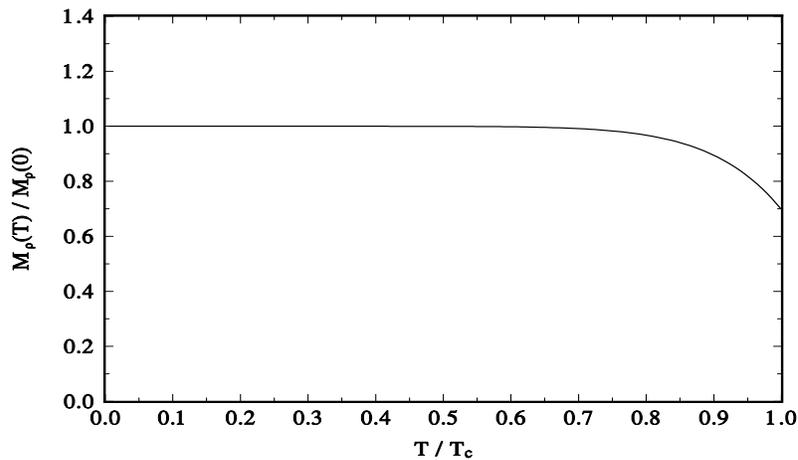}\caption{{\protect\small
			{Results from the FESR in the vector channel for the mass  of the $\rho$-meson, $M_\rho(T)/M_\rho(0)$,  Eq.(\ref{MASS}), as a function of $T/T_c$.}}}
	\label{figure11}
		\end{center}
\end{figure}

The $T$-dependence of the quark condensate was taken from LQCD \cite{LQCDqq2}-\cite{LQCDqq3}. Next, for the gluon condensate once again the LQCD results of \cite{LQCDGG} were used (see Fig.4). Finally, the remaining four-parameter space was mapped imposing as a constraint that the width, $\Gamma_\rho(T)$, should increase with increasing $T$, and that both  of the couplings, $f_\rho(T)$, and $s_0(T)$, should decrease with temperature. In this way the following thermal behaviour was obtained (for more details see \cite{rhoT})

\begin{equation}
\Gamma_\rho(T) = \frac{\Gamma_\rho(0)}{1- (T/T_c)^a}\;, \label{GAMMAf}
\end{equation}
where $a = 3$, and $T_c = 197 \,{\mbox{MeV}}$,
\begin{equation}
C_6\langle{\hat{O}_6}\rangle(T) = C_6\langle{\hat{O}_6}\rangle(0)\left[1- (T/T_q^*)^b\right]\;, \label{C6f}
\end{equation}
with $b=8$, and  $T_q^* = 187 \;{\mbox{MeV}}$, and
\begin{equation}
M_\rho(T) = M_\rho(0)\left[1- (T/T_M^*)^c\right]\;, \label{MASS}
\end{equation}
\begin{figure}[ht!]
	\begin{center}
	\includegraphics[height=3.2in, width=4.7in]{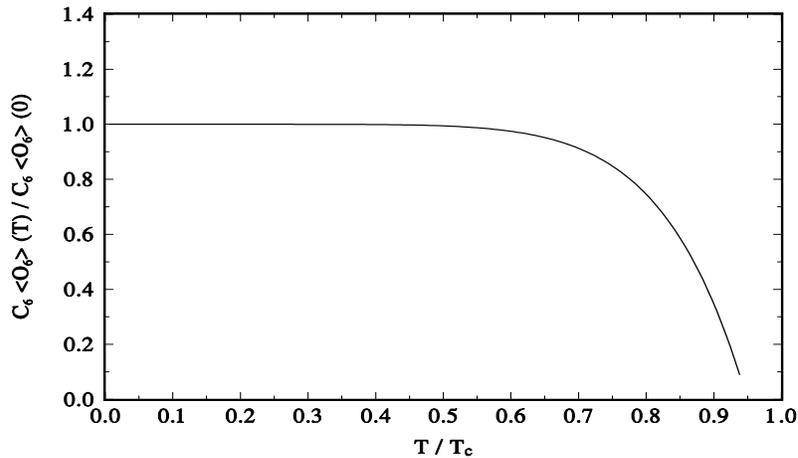}\caption{{\protect\small
			{The dimension $d \equiv 2 N = 6$ four-quark-condensate in the vector channel,  and in the vacuum saturation approximation, Eq.(\ref{C6f}), as a function of $T/T_c$.}}}
	\label{figure12}
		\end{center}
\end{figure}

where $c=10$, and $T_M^* = 222 \;{\mbox{MeV}}$, constrained to satisfy $T_M^* > T_c$. The slight $ 5 \%$ difference between $T_c$ and $T_q^*$ is well within the accuracy of the method. The remaining quantities are

\begin{equation}
s_0(T) = s_0(0) \left[ 1 - 0.5667 \;(T/T_c)^{11.38} - 4.347\; (T/T_c)^{68.41} \right], \label{S0TV}
\end{equation}

\begin{equation}
C_4\langle{\hat{O}_4}\rangle(T) = C_4\langle{\hat{O}_4}\rangle(0) \left[  1 - 1.65 \;(T/T_c)^{8.735} + 0.04967\; (T/T_c)^{0.7211}\right] \,. \label{C4O4T}
\end{equation}

\begin{equation}
f_\rho(T)/f_\rho(0) = 1 - 0.3901 \;(T/T_c)^{10.75} + 0.04155 \;(T/T_c)^{1.269} \label{frhoT}
\end{equation}

\begin{figure}[ht!]
	\centering
	\def\svgwidth{0.8\columnwidth}
	\includegraphics[height=3.5in, width=5.0in]{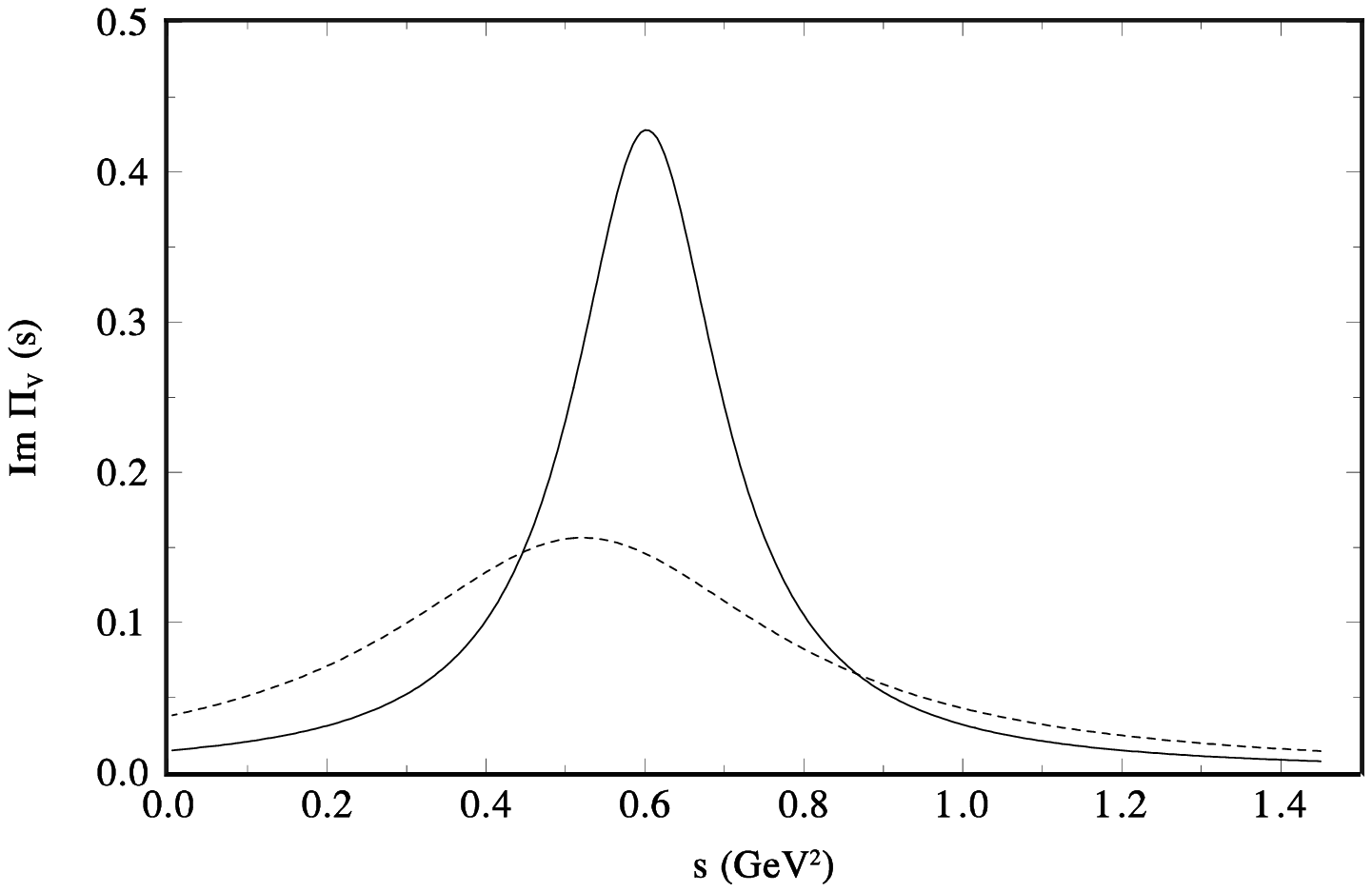}
	\caption{{\protect\small{The vector spectral function at $T=0$, Eq.(\ref{BW}) (solid curve), and at $T= 175 \;{\mbox{MeV}}$ (dotted curve) with thermal parameters given in Eqs. (\ref{GAMMAf}), (\ref{MASS}) and (\ref{frhoT}).}}}
	\label{Fig.13}
\end{figure}  
	
The behaviour of $s_0(T)$ is very similar to that in the axial-vector channel, Fig.6. The results for the coupling, $f_\rho(T)$, the width $\Gamma_\rho(T)$, the mass, $M_\rho(T)$,  and	$C_6\langle{\hat{O}_6}\rangle(T)$, all normalized to $T=0$, are shown in Figs. 9-12. Their behaviour is fully consistent with de-confinement taking place at at a critical temperature $T_c \simeq 190 - 200 \; {\mbox{MeV}}$. Of particular importance is the behaviour of the hadron mass. As shown in Fig.11, the hadron mass hardly changes with increasing $T$, particularly in relation to the behaviour of the hadronic width and coupling. A similar situation was found in the case of the heavy-light quark pseudo-scalar and vector-meson channels \cite{heavylight}. In fact, in the former channel, the hadron mass {\bf increases} by some 20\% at $T_c$, while the coupling vanishes and the width increases by a factor 1000. In the latter channel the mass {\bf decreases} by some 30\% while the coupling vanishes and the width increases by a factor 100. This should put to rest the ill-conceived idea that the $T$-behaviour of hadron masses is of any relevance to Physics at finite temperature. The hadronic vector spectral function is shown in Fig. 13 at $T=0$ (solid curve), and close to the critical temperature for de-confinement (dotted curve). The resonance broadening, together with the strong decrease of its peak value, as well as the decrease of the coupling, point to the disappearance of the $\rho$-meson from the spectrum. It should be pointed out that the correct parametrization of the $\rho$-spectral
function is as written in Eq.(\ref{BW}), as there is a misprint in \cite{rhoT}.  \\
\begin{figure}[ht!]
	\begin{center}
	\includegraphics[height=4.0in, width=4.9in]{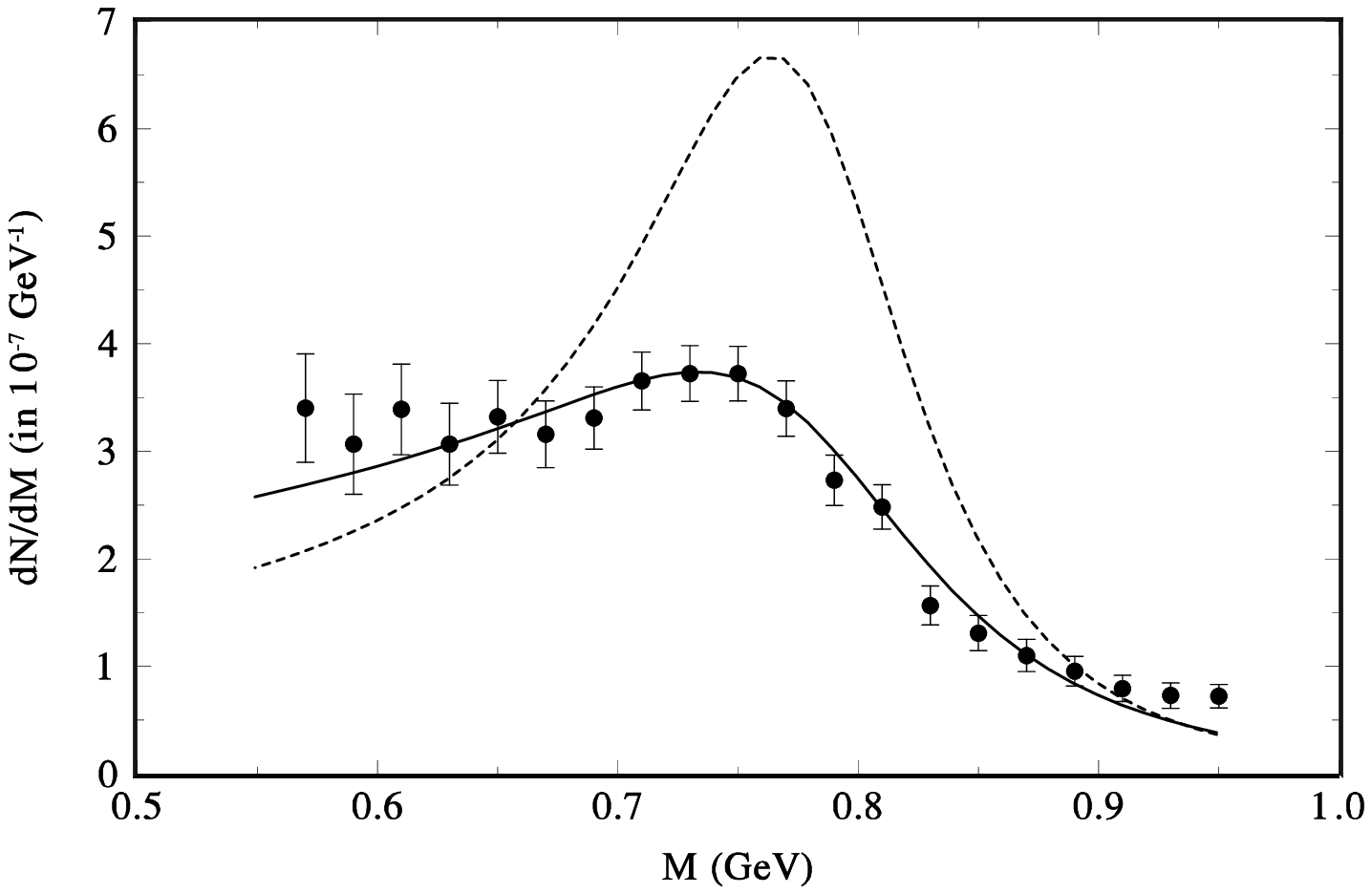}\caption{{\protect\small
	{The dimuon invariant mass distribution in In+In collisions in the region of the $\rho$-meson using Eq.(\ref{BW}) with  pre-determined values of thermal parameters from QCDSR, Eqs. (\ref{GAMMAf}), (\ref{MASS}), and (\ref{frhoT}) (solid curve). Dash curve is for all $\rho$-parameters independent of $T$. The predicted resonance broadening, as well as the flattening of the invariant mass distribution near the peak are clearly observed.  Data is from \cite{NA601}-\cite{NA605}. Results are for $\mu =0$. Finite chemical potential results change slightly in off-peak regions (see \cite{dimuonAD}) .}}}
\label{figure14}	
	\end{center}
\end{figure}

To complete this section we describe how to obtain the dimuon production rate in heavy ion collisions at high energy, in particular for the case of In + In (at 158 A GeV) into $\mu^+ \mu^-$, as measured by CERN NA60 Collaboration \cite{NA601}-\cite{NA605}. The  issues in dimuon production were discussed long ago in \cite{dimuon1}-\cite{JWAC}. In particular, in \cite{JWAC} a detailed analysis of this process, using  
Bjorken's scaling solution for longitudinal hydrodynamic expansion \cite{JB}, was discussed. Predictions for the dimuon production rate were also made in \cite{JWAC} assuming the pion form factor, entering the production rate, to be dominated by the $\rho$-meson with parameters strictly from $T=0$ in Eq.(\ref{BW}). A proposal to use instead a $T$-dependent hadronic width in the $\rho$-meson spectral function was  first made in \cite{dimuonDL}, and re-discovered several years later in \cite{EIoffeK1}-\cite{EIoffeK2}. It must be mentioned that at the time of this proposal \cite{dimuonDL}, this idea was truly innovative. It was shown in \cite{dimuonDL}, using some primitive model for $\Gamma_\rho(T)$, that there would be important, detectable changes in the production rate, such as, e.g. a flattening of the rate around the $\rho$-peak, together with a broadening of this peak. This prediction was made in 1991, way before any experimental data were available, and $T$-dependent hadron widths hardly used. By the time data became available, the proposal had been forgotten, but recent experimental results fully confirmed the idea of a $T$-dependent $\rho$-meson width, and the prediction of a flattening rate with increasing $T$, as shown next.\\

The dimuon production rate involves several kinematical and dynamical factors (see \cite{dimuonDL}, \cite{JWAC}), including the $\rho$-meson hadronic spectral function. Recently, in a reanalysis of this process \cite{dimuonAD}, the latter was parametrized as in Eq.(\ref{BW}) but with $T$-dependent parameters, given in Eqs.(\ref{GAMMAf}), (\ref{MASS}), and (\ref{frhoT}). Furthermore, in addition to the temperature, it turns out that the chemical potential (density), $\mu$, needs to be introduced. This topic  will be discussed in Section 8 following \cite{mu} where a QCDSR analysis at finite $\mu$ was first proposed. The parameter-free prediction of the dimuon invariant mass distribution is shown in Fig. 14 (solid curve), together with the NA60 data \cite{NA601}-\cite{NA605}, and the prediction using a $T$ independent spectral function (dash curve). The predicted resonance broadening, essentially from Eq.(\ref{GAMMAf}), as well as the flattening of the spectrum around the peak are fully confirmed. It must be pointed out that this determination is only valid in the region around the $\rho$-peak. At lower, as well as at higher energies this approximation breaks down, and the dynamics would involve a plethora of processes hardly describable in QCD. Intermediate energy models of various kinds have been invoked to account for the data in those regions, with varying degrees of success (for a recent review see \cite{Rapp}).

\begin{figure}[ht!]
	\centering
	\def\svgwidth{0.8\columnwidth}
	\includegraphics[height=3.5in, width=5.0in]{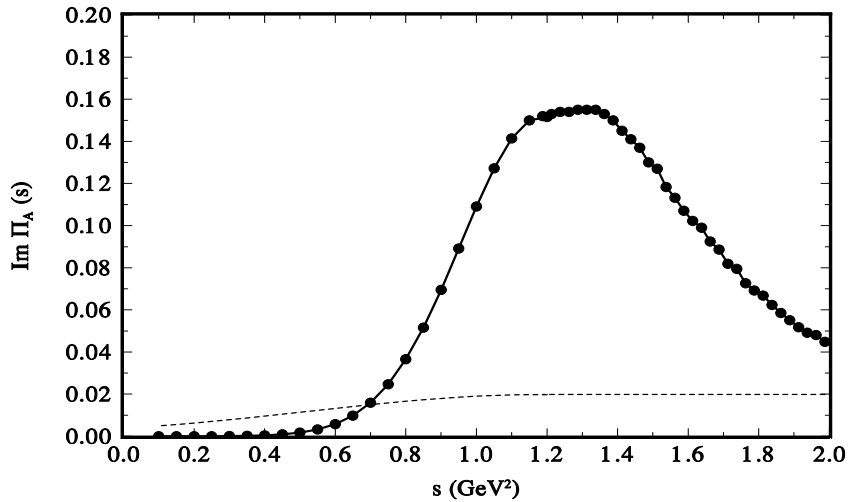}
	\caption{{\protect\small{Solid curve is the axial-vector ($a_1$-resonance) spectral function at $T=0$ fitted to the ALEPH data \cite{ALEPH3}, shown with  error bars the size of the data points. Dotted curve is the spectral function at  $T= 175 \;{\mbox{MeV}}$ with thermal parameters given in Eq. (\ref{YT}).}}}
\end{figure} 

\section{Weinberg sum rules and chiral mixing at finite temperature}

The two Weinberg sum rules (WSR) (at $T=0$)) \cite{WSR0} were first derived in the framework of chiral $SU(2) \times SU(2)$ symmetry and current algebra, and read

\begin{equation}
W_1 \equiv 
\int\limits_{0}^{\infty} ds \,\frac{1}{\pi}\, [{\mbox{Im}} \Pi_V(s) - {\mbox{Im}} \Pi_A(s) ] = 2 \, f_\pi^2\;, \label{WSR1}
\end{equation}

\begin{equation}
W_2 \equiv 
\int\limits_{0}^{\infty} ds \,s \,\frac{1}{\pi}\, [{\mbox{Im}} \Pi_V(s) - {\mbox{Im}} \Pi_A(s) ] =0 \;. \label{WSR2}
\end{equation}
 
Given that both the vector and the axial-vector spectral functions enter in the WSR,  unfortunately we need to change the previous normalization of the vector correlator, Eq.(\ref{Pi}), to turn it equal to that of the axial-vector one, Eq.(\ref{Pi0Axial}), i.e. we choose

\begin{equation}
{\mbox{Im}} \,\Pi_{V}(q^2) = {\mbox{Im}} \,\Pi_A(q^2) = \frac{1}{4\pi}\left[ 1 + {\cal{O}} \left( \alpha_s(q^2)\right)\right]\,. \label{VANORM}
\end{equation}

In the framework of perturbative QCD (PQCD), where both spectral functions have the same  asymptotic behaviour (in the chiral limit), these WSR become effectively QCD finite energy sum rules (FESR)

\begin{equation}
W_{n+1}(s_0) \equiv \int\limits_{0}^{s_0} ds \, s^n\, \frac{1}{\pi}\, [{\mbox{Im}} \Pi_V(s) - {\mbox{Im}} \Pi_A(s) ] = 2 \, f_\pi^2\, \delta_{n 0}\;, \label{WSR}
\end{equation}

where $s_0 \simeq 1-3 \, {\mbox{GeV}^2}$, is the squared energy beyond which QCD is valid, and both sum rules have been combined. This result also follows from Cauchy's theorem in the complex s-plane, together with the assumption of quark-hadron duality, Eq.(\ref{FESR}). The latter is not expected to hold in the resonance region where QCD is not valid on the positive real $s$-axis. This leads to duality violations (DV), first identified long ago in \cite{Shankar}, and currently a controversial subject of active  research \cite{pinched11}-\cite{pinched12}, \cite{pinched2}. In relation to the WSR, it was pointed out long ago \cite{DV2} that these sum rules were hardly satisfied by saturating them with the ALEPH data on hadronic $\tau$-lepton decays \cite{ALEPH1}-\cite{ALEPH3}. This was, and still can be interpreted as a signal for DV. A proposal was made in \cite{DV2} (see also \cite{DV3}) to introduce the non-trivial kernel $P(s)$ in Eq. (\ref{FESR}), leading to

\begin{equation}
WP(s_0) \equiv
\int\limits_{0}^{s_0} ds \left( 1 - \frac{s}{s_0}\right) \frac{1}{\pi}\; [{\mbox{Im}} \Pi_V(s) - {\mbox{Im}} \Pi_A(s) ]
= 2\;  f_\pi^2 \,.\label{WP}
\end{equation}

This expression is fully satisfied \cite{pinched11}-\cite{pinched12},\cite{DV2}, thus validating the procedure. It turns out that this is also the case in other sum rules \cite{DV2}, i.e. pinched kernels quench or even eliminate DV.\\
Turning to the WSR at finite temperature \cite{WSRT}, all parameters in the vector channel have been already determined in Eqs.(\ref{GAMMAf}), (\ref{MASS}), and (\ref{S0TV}). The axial-vector channel parameters  at $T=0$ require a slight update, as they were obtained in the squared energy region below $s \simeq 1.5 \, {\mbox{GeV}}^2$. Going above this value, and up to $s \simeq 2.0 \, {\mbox{GeV}^2}$, the resonance hadronic spectral function at $T=0$ fitted to the ALEPH  $\tau$-decay data is \cite{WSRT}

\begin{equation}
\frac{1}{\pi} {\mbox{Im}} \Pi_A(s)|_{a_1} = C\, f_{a_1} \exp\left[-\left(\frac{s- M_{a_1}^2}{\Gamma^2_{a_1}}\right)^2\right] 
\;\;\;\;\;\;\;\;\;\;\;\;\;\;\;\;\;\;
(0 \leq s \leq 1.2\, {\mbox{GeV}}^2)\;,\label{A11}
\end{equation}

\begin{equation}
\frac{1}{\pi} {\mbox{Im}} \Pi_A(s)|_{a_1} = C\, f_{a_1} \exp\left[-\left(\frac{1.2\; {\mbox{GeV}}^2 - M_{a_1}^2}{\Gamma^2_{a_1}}\right)^2\right] 
\;\;\;\;\;\;\;\;
(1.2\, {\mbox{GeV}}^2 \leq s \leq  1.45\,{\mbox{GeV}}^2)\;,\label{A12}
\end{equation}

\begin{equation}
\frac{1}{\pi} {\mbox{Im}} \Pi_A(s)|_{a_1} = C\, f_{a_1} \exp\left[-\left(\frac{s- M_{a_1}^2}{\Gamma^2_{a_1}}\right)^2\right] 
\;\;\;\;\;\;\;\;\;\;\;\;\;\;\;\;\;\;\;\;\;\;
(1.45\, {\mbox{GeV}}^2 \leq s \leq M_\tau^2)\;,\label{A13}
\end{equation}

where $M_{a_1} = 1.0891 \; {\mbox{GeV}}$, $\Gamma_{a_1} = 568.78\; {\mbox{MeV}}$, $C= 0.662$ and $f_{a_1}= 0.073$ (the latter two parameters were split to facilitate comparison between $f_{a_1}$ and $f_{\rho}$, for readers used to zero-width resonance saturation of the WSR). 
The results from the FESR for the thermal parameters can be written as

\begin{equation}
\frac{Y(T)}{Y(0)} \,=\, 1 \, + \, a_1\, \left(\frac{T}{T_c}\right)^{b_1}\, + \,a_2 \, \left(\frac{T}{T_c}\right)^{b_1} \;, \label{YT}
\end{equation}

where the various coefficients are given in Table 1. 
\begin{table}[hb]
	\begin{center}
		\begin{tabular}{ccccc}
			\hline 
			\multicolumn{4}{r}
			{Coefficients in Eq. (54)} \\
			\cline{2-5}
			\noalign{\smallskip}
			Parameter  & $a_1$ &$a_2$  & $b_1$ &$b_2$   \\
			\hline
			\noalign{\smallskip}
			$s_0(T)$ & - 28.5 & -0.6689 & 35.60 & 3.93  \\
			$f_\pi(T)$ &- 0.2924 & - 0.7557 & 73.43 & 11.08   \\
			$f_{a_1}(T)$ & - 19.34 & 14.27 & 7.716 & 6.153 \\
			$\Gamma_{a_1}(T)$&2.323 & 1.207 & 20.24 & 7.869\\
			\hline 
		\end{tabular}
		\caption{\footnotesize{The values of the coefficients entering Eq. (54).}}
	\end{center}
	
\end{table}
\begin{figure}[ht!]
	\centering
	\def\svgwidth{0.8\columnwidth}
	\includegraphics[height=3.2in, width=4.5in]{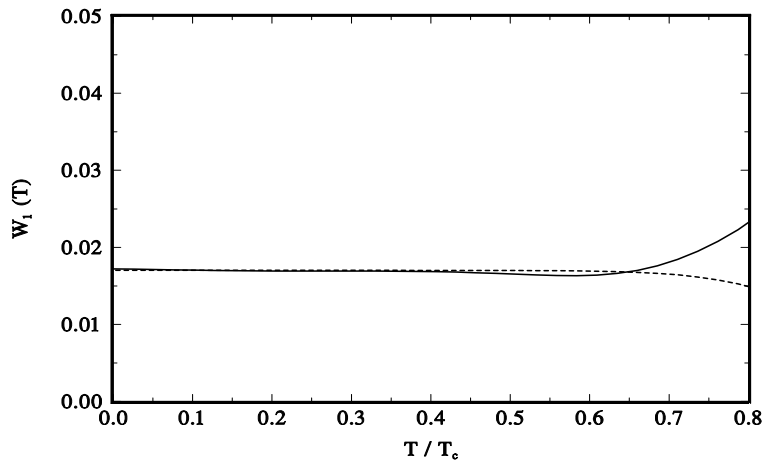}
	\caption{{\protect\small{The first WSR, Eq.(\ref {WSR1}),  at finite T. Solid (dash) line is the left (right) hand side of Eq.(\ref{WSR1}). The divergence at high $T$ is caused by the asymmetric hadronic {\it{scattering}} contribution in the space-like region ($q^2 <0$), which disappears at de-confinement ($T=T_c$).}}} 
\end{figure}

The $a_1$ mass hardly changes with temperature, so that it was kept constant. This behaviour can be ascribed to the very large width of the $a_1$ resonance.\\
A comparison of the vector spectral function at $T=0$, and at $T = 175 \, {\mbox{MeV}}$ is shown in Fig.13. Resonance broadening, with a slight decrease of the mass is clearly seen. In the axial-vector case the spectral function is shown in Fig.15, where the solid curve is the fit to the ALEPH data at $T=0$, and the dotted curve corresponds to $T = 175\; {\mbox{MeV}}$. At such temperature there is no trace of the $a_1$.\\

Proceeding to the WSR at finite $T$, the first obvious thing to notice is the dramatic difference between the vector and the axial-vector spectral functions. These spectral functions have very different evolution with increasing temperature, for the obvious reason that they are already so different at $T=0$, perhaps with the exception that $s_0(0)$ is the same in both channels. With increasing $T$, the parameters of each channel evolve independently, thus keeping both spectral functions distinct. Eventually, 
this asymmetry  is expected to vanish at de-confinement, when the $\rho$ and the $a_1$ mesons disappear from the spectrum. This implies no chiral-mixing at any temperature, except obviously at $T\simeq T_c$. In addition to these differences there is an additional asymmetry due to the hadronic (pionic) scattering term, present in the vector channel at the leading, one-loop level, and strongly two-loop suppressed in the axial-vector case. This is manifest very close to the critical temperature, where  this term is important, as it increases quadratically with temperature. This can be appreciated in Fig.16, which shows the $T$-dependence of the first WSR, $W_1(T)$, Eq.(\ref{WSR1}). The behaviour of the pinched WSR, $WP(T)$,  Eq.(\ref{WP}), is essentially the same, except close to $T_c$ where the scattering term causes $WP(T)$ to decrease, rather than increase, slightly. In both cases, the scattering term disappears at $T=T_c$, as the pions would have melted. To be more specific let us consider the vector and axial-vector correlators, Eqs. (\ref{VVcorrelator}) and (\ref{AAcorrelator}), respectively. In a thermal bath, and in the hadronic representation one has (schematically)

\begin{equation}
\Pi_{\mu\nu}|_{V} = \langle\pi| V_\mu(0) V_\nu(x)| \pi \rangle
= \langle\pi| V_\mu(0)|\pi\rangle \langle\pi| V_\nu(x)| \pi \rangle + \langle\pi \pi| V_\mu(0)|\pi \pi\rangle \langle\pi\pi| V_\nu(x)| \pi \pi\rangle + \cdots \,,\label{VVpi}
\end{equation}

\begin{equation}
\Pi_{\mu\nu}|_{A} = \langle\pi| A_\mu(0) A_\nu(x)| \pi \rangle
= \langle\pi| A_\mu(0)|0\rangle \langle 0| A_\nu(x)| \pi \rangle + \langle \pi \pi \pi| A_\mu(0)|0 \rangle \langle 0| A_\nu(x)| \pi \pi \pi\rangle + \cdots \,. \label{AApi}
\end{equation}

To the extent that Isospin and G-parity remain valid symmetries at finite temperature, the chiral asymmetry is manifest, to wit. The leading term in the vector channel is the two-pion one-loop, and in the axial-vector channel it is the tree-level pion to-vacuum term ($f_\pi$), followed by a highly phase-space suppressed three-pion two-loop. In other words, the matrix element $\langle\pi| A_\mu(0)|\pi\rangle$, invoked by chiral-mixing proposers \cite{Dey}, vanishes identically at leading order. The correct matrix element, beyond the pion pole, is the phase-space suppressed second term in Eq.(\ref{AApi}). In principle, this term  could have a resonant sub-channel contribution from the matrix element $\langle \rho\,\pi| A_\mu(0)|0 \rangle$, which again is phase space suppressed  (see results from \cite{EdRP1}-\cite{EdRP2} which can be easily adapted to this channel).
An alternative argument clearly showing the non-existence of chiral mixing at finite $T$ is based on the chiral Lagrangian to leading order \cite{Mallik}, with all terms respecting Isospin and G-parity.

\section{Temperature dependence of the up- down- quark mass}

In this section we discuss a recent determination of the thermal dependence of the up- down-quark mass \cite{mqT}. At $T=0$ the values of the light-quark masses are determined from QCD sum rules usually involving the correlator of the axial-vector divergences \cite{QCDBook},\cite{quarkmassR1}-\cite{quarkmassR4}

\begin{equation}
\psi_{5}(q^{2})= i\int d^{4}x \,  e^{i q x} \, \langle 0| T(\partial^\mu A_\mu(x) \, \partial^\nu A^\dagger_\nu(0))|0\rangle , \label{psi5}
\end{equation}

with

\begin{equation}
\partial^\mu A_\mu(x) = \overline{m}_{ud} : \overline{d}(x)\, i \,\gamma_5\, u(x): \;, \label{DmuAmu}
\end{equation}

and the definition

\begin{equation}
\overline{m}_{ud} \equiv (\overline{m}_u+\overline{m}_d) \, \simeq \, 10 \,{\mbox{MeV}} \,, \label{mud}
\end{equation}

where $\overline{m}_{u,d}$ are the running quark masses in the $\overline{MS}$-scheme at a scale $\mu = 2  \, {\mbox{GeV}}$ \cite{QCDBook}, \cite{quarkmassR1}-\cite{quarkmassR4}, \cite{FLAG} and $u(x)$, $d(x)$, the corresponding quark fields. As usual, the relation between the QCD and the hadronic representation of current correlators is obtained by invoking Cauchy's theorem in the complex square-energy plane, Fig.2, which leads to the FESR 

\begin{equation}
\int_{\mathrm{0}}^{s_0} ds\, \frac{1}{\pi}\, Im \,\psi_5(s)|_{HAD}  = -\frac{1}{2 \pi i}  \oint_{C(|s_0|) }\, ds  \,\psi_5(s)|_{QCD} \, , \label{FESRD1}
\end{equation}

\begin{equation}
 \int_{\mathrm{0}}^{s_0} \frac{ds}{s}\, \frac{1}{\pi}\, Im \,\psi_5(s)|_{HAD}  + \frac{1}{2 \pi i}  \oint_{C(|s_0|) }\, \frac{ds}{s}  \,\psi_5(s)|_{QCD}  = \psi_5(0) \;,\label{FESRD2}
\end{equation}

where $\psi_5(0)$ is the residue of the pole generated by the denominator in Eq.(\ref{FESRD2}), i.e.

\begin{equation}
\psi_5(0)= {\mbox{Residue}}\, [\psi_5(s)/s]_{s=0} \,.\label{psi50}
\end{equation}

\begin{figure}[ht!]
	\centering
	\def\svgwidth{0.8\columnwidth}
	\includegraphics[height=3.0in, width=4.5in]{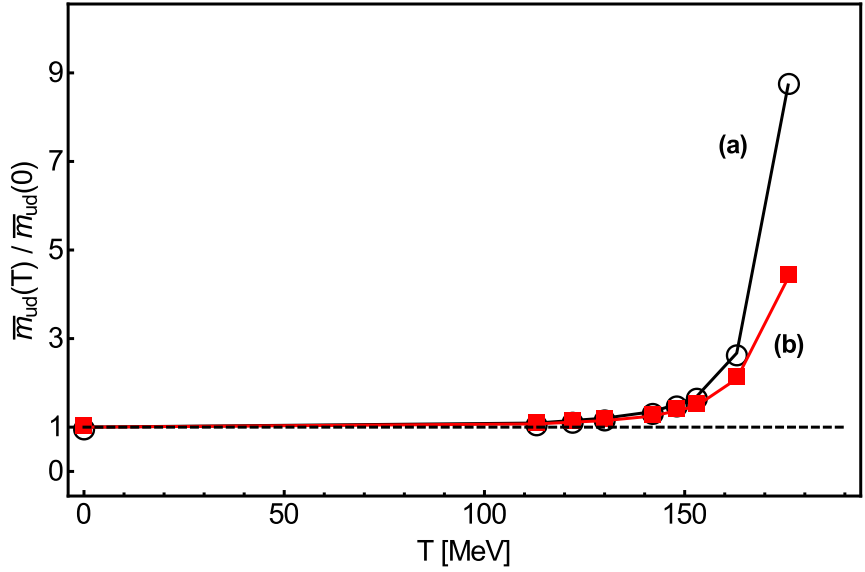}
	\caption{{\protect\small{The ratio of the quark masses $\overline{m}_{ud}(T)/\overline{m}_{ud}(0)$ as a function of $T$ from the FESR Eqs.(\ref{FESRD1T})-(\ref{FESRD2T}). Curve (a) is for a $T$-dependent pion mass from \cite{MPIT}, and curve (b) is for a constant pion mass.}}} 
\end{figure}

The radius of the contour, $s_0$, in Fig.2 is large enough for QCD to be valid on the circle. Information on the hadronic spectral function on the left hand side of Eq.(\ref{FESRD1}) allows to determine the quark masses entering the contour integral. Current precision determinations of quark masses \cite{QCDBook}, \cite{quarkmassR1}-\cite{quarkmassR4} require the introduction of integration kernels on both sides of Eq.(\ref{FESRD1}). These kernels are used to enhance or quench hadronic contributions, depending on the integration region, and on the quality of the hadronic information available. They also deal with the issue of potential quark-hadron duality violations, as QCD is not valid on the positive real axis in the resonance region. This will be of no concern here, as we are going to determine only  ratios, e.g. $\overline{m}_{ud}(T)/\overline{m}_{ud}(0)$, to leading order in the hadronic and the QCD sectors. To this order, the QCD expression of the pseudoscalar correlator, Eq.(\ref{psi5}), is

\begin{equation}
\psi_5(q^2)|_{QCD} =  \overline{m}_{ud}^2 
\left\{- \frac{3}{8 \,\pi^2} q^2  \ln\left(\frac{- q^2}{\mu^2}\right)
 +  \frac{\overline{m}_{ud} \;\langle \bar{q}q \rangle}{q^2} - \frac{1}{8\,  q^2} \;\langle \frac{\alpha_s}{\pi} G^2 \rangle + {\cal{O}} \left( \frac{O_6}{q^4} \right)\right\} ,\label{psi5QCD0}
\end{equation} 

where $\langle \bar{q} q \rangle = (- 267 \pm 5 \, {\mbox{MeV}})^3$ from \cite{GMOR},  and $\langle \frac{\alpha_s}{\pi} G^2  \rangle = 0.017 \pm 0.012 \, {\mbox{GeV}}^4$ from \cite{GG}.
The  gluon- and quark-condensate contributions in Eq.(\ref{psi5QCD0}) are, respectively, one and two orders of magnitude smaller than the leading perturbative QCD term. Furthermore, at finite temperature both condensates decrease with increasing $T$, so that they can be safely ignored in the sequel.\\
\begin{figure}[ht]
  \begin{center}
    \includegraphics[height=2.7 in,width=4.5 in]{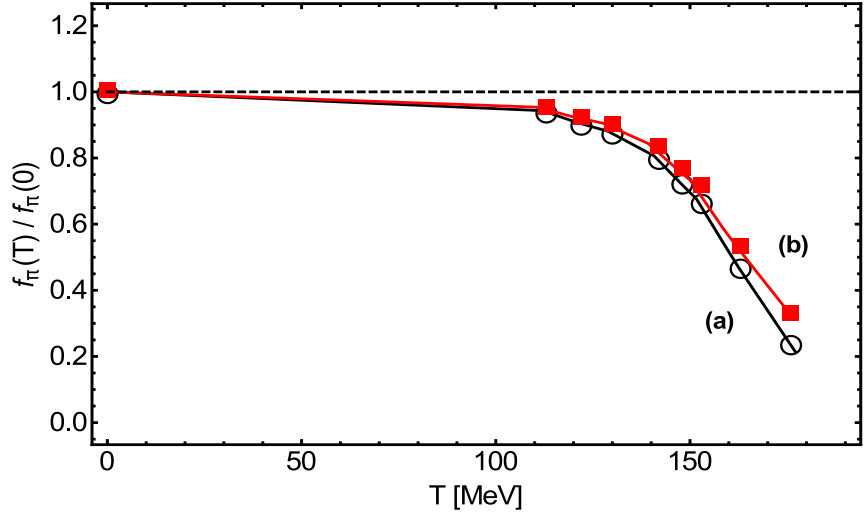}
    \caption{{\protect \small{The ratio of the pion decay constant $f_\pi(T)/f_\pi(0)$ as a function of $T$ from the FESR Eqs.(\ref{FESRD1T})-(\ref{FESRD2T}). Curve (a) is for a $T$-dependent pion mass from \cite{MPIT}, and curve (b) is for a constant pion mass.}}}
  \end{center}
  \end{figure}

The QCD spectral function in the time-like region, at finite $T$, obtained from the Dolan-Jackiw formalism \cite{DJ} in the rest frame of the medium $(q^2 = \omega^2 - |\bf{q}|^2 \rightarrow \omega^2)$ is given by

\begin{equation}
{\mbox{Im}}\; \psi_5(q^2,T)|_{QCD}= \frac{3}{8 \, \pi} \overline{m}_{ud}^2 (T) \; \omega^2  \left[ 1 \,-\, 2 \, n_F(\omega/ 2T) \right]\,. \label{psi5QCDT}
\end{equation}

The QCD scattering term, present in the axial-vector correlator, Eq.(\ref{Pi0A-T}), is absent in $\psi_5(q^2,T)$, due to the overall factor of $q^2$ in the first term in Eq.(\ref{psi5QCD0}). This factor prevents the appearance of the delta function, $\delta(\omega^2)$, in Eq.(\ref{Pi0A-T}). In the hadronic sector the scattering term is due to a phase-space suppressed two-loop three-pion contribution, which is negligible in comparison with the pion-pole term

\begin{equation}
{\mbox{Im}}\; \psi_5(q^2,T)_{HAD}=
2 \, \pi\,f_\pi^2(T) \, M_\pi^4(T) \; \delta(q^2-M_\pi^2)\,. \label{Pionpole}
\end{equation}

The two FESR, Eqs.(\ref{FESRD1})-(\ref{FESRD2}), at finite $T$ become

\begin{equation}
2 f_\pi^2(T) M_\pi^4(T) = \frac{3 \,\overline{m}_{ud}^2(T)}{8 \pi^2}\int_0^{s_0(T)} s \left[1-2 n_F \left(\frac{\sqrt{s}}{2 T}\right)\right]ds , \label{FESRD1T}
\end{equation}

\begin{equation}
2 \,f_\pi^2(T) \, M_\pi^2(T) = - 2 \,\overline{m}_{ud}(T) \,\langle\bar{q} q\rangle(T) \, + \, \frac{3}{8 \pi^2} \,\overline{m}_{ud}^2(T) 
\int_0^{s_0(T)} \left[1 - 2 n_F \left(\frac{\sqrt{s}}{2 T}\right)\right] \,ds. \label{FESRD2T}
\end{equation}

Equation (\ref{FESRD2T}) is the thermal Gell-Mann-Oakes-Renner relation incorporating a higher order QCD quark-mass correction, ${\cal{O}} (m^2_{ud})$. While at $T=0$ this correction is normally neglected \cite{GMOR}, at finite temperature this cannot be done, as it is of the same order in the quark mass as the right-hand-side of Eq.(\ref{FESRD1T}).\\

As discussed previously in Section 3, the thermal quark condensate (signalling chiral-symmetry restoration) and $s_0(T)$ (signalling deconfinement) are related through

\begin{equation}
\frac{s_0(T)}{s_0(0)}\simeq \left[\frac{\langle \bar{q} q \rangle(T)}{\langle \bar{q} q \rangle(0)}\right]^{2/3}. \label{s0qq}
\end{equation}

Further support for this relation is provided by LQCD results \cite{LQCD2a}-\cite{LQCD2b}. One does not expect this relation to be valid very close to the critical temperature, $T_c$, as  the thermal quark condensate, for finite quark masses is non-vanishing close to $T_c$.  With $s_0(T)/s_0(0)$ as input in the FESR, Eqs. (\ref{FESRD1T})-(\ref{FESRD2T}), together with  LQCD results for $\langle \bar{q} q\rangle(T)$ for finite quark masses \cite{LQCD}, and independent determinations of $M_\pi(T)$ \cite{MPIT},   the ratios $\overline{m}_{ud}(T)/\overline{m}_{ud}(0)$ and $f_\pi(T)/f_\pi(0)$ were obtained in \cite{mqT}. The results are shown in Figs.17 and 18. The quark mass remains constant up to $T \simeq 150 \, {\mbox{MeV}}$, and increases sharply thereafter. As expected from the discussion on chiral-symmetry in Section 1, leading to Eq. (\ref{Mpi2}), the quark mass is intimately related to the pion mass. The behaviour of the quark mass is also consistent with the expectation that at deconfinement free light quarks would acquire a much higher constituent mass. Figure 18 shows the thermal behaviour of $f_\pi$, which is fully consistent with the expectation from chiral-symmetry, Eq.(\ref{fpi2}), i.e. that $f_\pi(T)$ is independent of $M_\pi(T)$, and $f_\pi(T) \propto \langle \bar{q} q\rangle(T)$.
\section{Quarkonium at finite temperature and its survival at/beyond $T_c$}

In 1986 Matsui and Satz \cite{MS}, invoking colour screening in charmonium, concluded that this effect would prevent  binding in the de-confined interior of the interaction region in heavy-ion collisions. This scenario became an undisputed mantra for more than two decades, until 2010  when it was shown \cite{ccbar1} that thermal QCD sum rules clearly predict the survival of charmonium ($J/\psi$) at and beyond $T_c$. Subsequently, this was supported by an analysis of scalar and pseudoscalar charmonium states \cite{ccbar2}, and pseudoscalar and vector bottonium states \cite{bbar}, all behaving similarly to the $J/\psi$.  The results for bottonium were in qualitative agreement with LQCD simulations \cite{LQCD_QQbar1}-\cite{LQCD_QQbar2}. An interesting aspect of the latter is the result for the widths. In fact, the qualitative temperature behaviour of hadronic widths from LQCD agrees with that from QCDSR. This is reassuring given that these two approaches employ very different parameters to describe deconfinement. Recent work \cite{Polyakov} shows that these two parameters, $s_0(T)$ for the thermal QCDSR, and the Polyakov thermal loop for LQCD, are in fact related as they provide the same information on deconfinement.\\
\begin{figure}[ht!]
\begin{center}
\includegraphics[height=3.2 in,width=4.5 in]{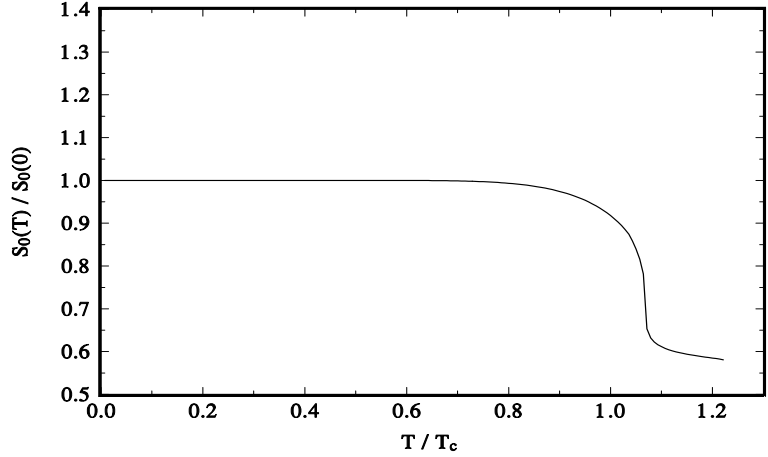}
\caption{\protect\small{The  ratio $s_0(T)/s_0(0)$  as a function of $T/T_c$ for the $J/\psi$ channel, from thermal Hilbert moment QCD sum rules.}}
\end{center}
\end{figure}

We proceed to discuss the thermal behaviour of charmonium in the vector channel \cite{ccbar1}, i.e. the $J/\psi$ state. The case of scalar and pseudoscalar charmonium \cite{ccbar2}, as well as bottonium \cite{bbar}, follows along similar lines, so the reader is referred to the original papers for details. The vector current correlator is given by Eq.(\ref{VVcorrelator}), with the obvious replacement of the light- by the heavy-quark fields in the vector current, $V_\mu(x) = : \bar{Q}(x) \gamma_\mu Q(x):$, where $Q(x)$ is the charm-quark field. A straightforward calculation in the time-like region, to leading order in PQCD, gives 

\begin{equation}
\frac{1}{\pi}\, Im \,\Pi^a(q^2,T) =  \frac{3}{16 \pi^2}\;\int_{-v}^{v}\;dx \;(1-x^2) \left[1 - n_F\left(\frac{|\mathbf{q}| x + \omega}{2 T}\right)
- n_F\left( \frac{|\mathbf{q}| x -\omega}{2 T} \right)\right] \;,\label{PIVQ}
\end{equation}

where $ v^2 = 1 - 4 m_Q^2/q^2$, $m_Q$ is the heavy quark mass, $q^2 = \omega^2 - \mathbf{q}^2 \geq 4 m_Q^2$, and  $n_F(z)$  is the Fermi thermal function. In the rest frame of the thermal bath, $|\mathbf{q}| \rightarrow 0$, the above result reduces to

\begin{equation}
\frac{1}{\pi}\, Im \,\Pi^a(\omega,T) =  \frac{1}{8 \pi^2}\; v (3 - v^2) \left[1 - 2 n_F (\omega/2 T)\right] \; \theta(\omega - 2 m_Q)
\;.\label{PIVQq0}
\end{equation}

\begin{figure}[ht!]
\begin{center}
\includegraphics[height=3.2 in,width=4.5 in]{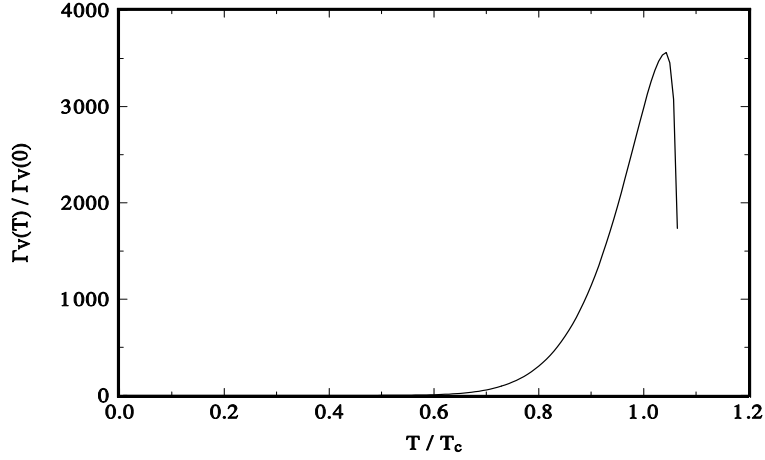}
\caption{{\protect\small The  ratio $\Gamma_V(T)/\Gamma_V(0)$  as a function of $T/T_c$ for the $J/\psi$ channel, from thermal Hilbert moment QCD sum rules}}
\end{center} 
\end{figure}

The quark mass is assumed independent of $T$, which is a good approximation for temperatures below 200 MeV \cite{mQ}. In the space-like region the QCD scattering term, Eq.(\ref{Pi-V}), needs to be re-evaluated to take the quark mass into account This gives 

\begin{equation}
\frac{1}{\pi}\, Im \,\Pi^s(\omega,T) =  \frac{2}{ \pi^2}\; m_Q^2\; \delta(\omega^2)\left[   n_F \left(\frac{m_Q}{T}\right)\;+ \frac{2\, T^2}{m_Q^2} \;\int_{ m_Q/T }^{\infty} y \,n_F (y)\, dy \;\right]
\;.\label{Pi-VQ}
\end{equation}

\begin{figure}[ht!]
\begin{center}
\includegraphics[height=3.2 in,width=4.5 in]{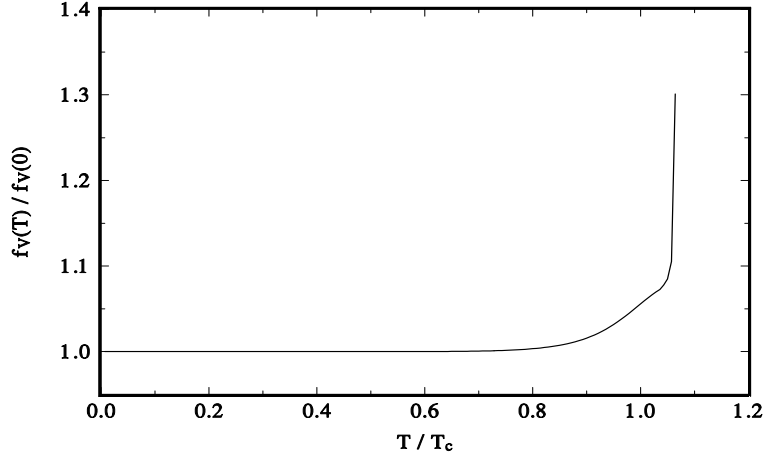}
\caption{\protect\small{The  ratio $f_V(T)/f_V(0)$ as a function of $T/T_c$ for the $J/\psi$ channel, from thermal Hilbert moment QCD sum rules}}
\end{center}
\end{figure}
In the hadronic sector the spectral function is given by the ground-state pole, $J/\psi$, followed by PQCD
\begin{figure}[ht!]
\begin{center}
\includegraphics[height=3.2 in,width=4.5 in]{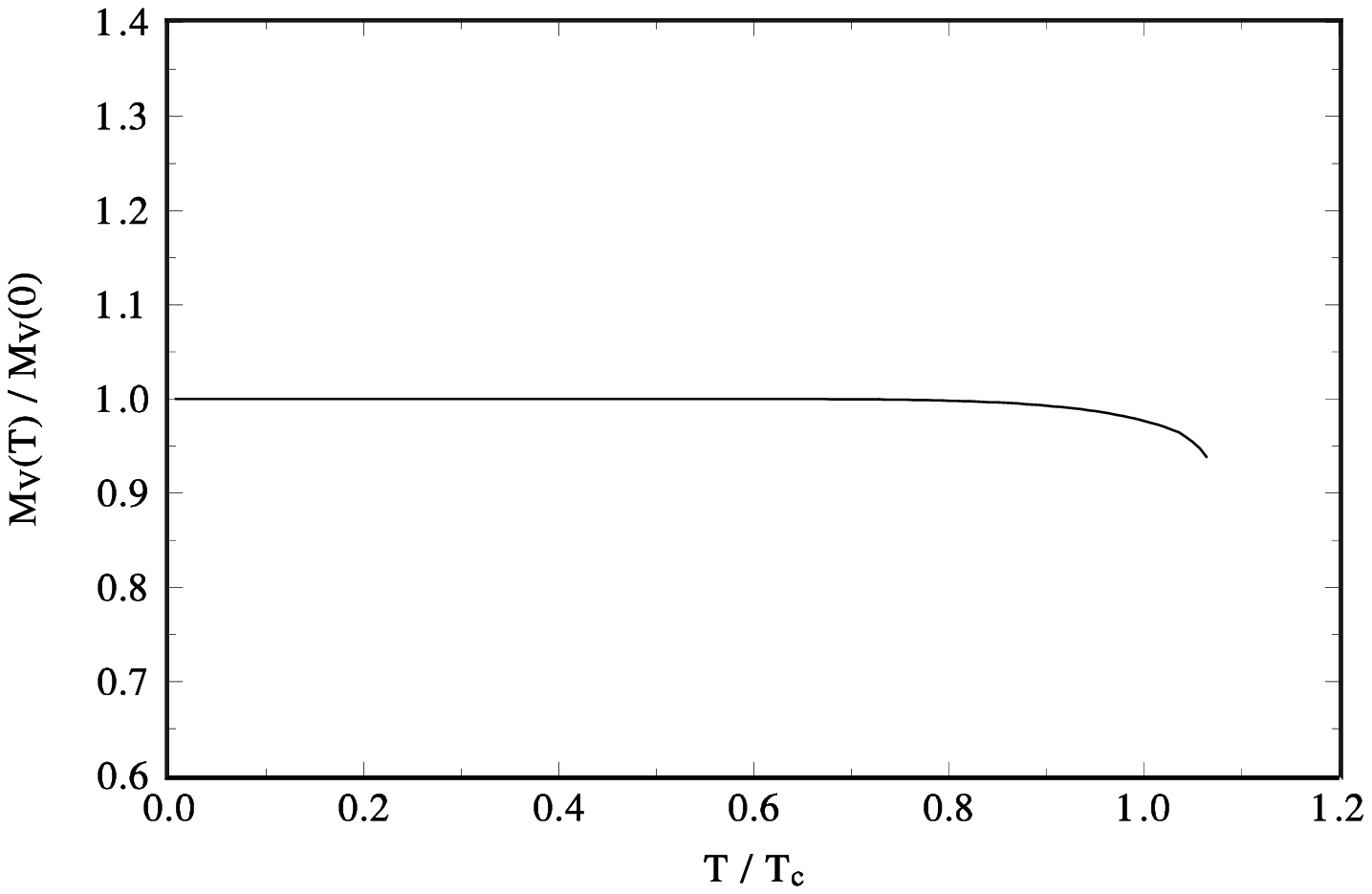}
\caption{\protect\small{The  ratio $M_V(T)/M_V(0)$ as a function of $T/T_c$ for the $J/\psi$ channel, from thermal Hilbert moment QCD sum rules. This ratio is basically the same in zero-width as in finite width.}}
\end{center}
\end{figure} 
\begin{equation}
\frac{1}{\pi}\, Im \,\Pi(s,T)|_{HAD} =   2 \, f_V^2(T)  \, \delta(s - M_V^2(T)) \,+  \frac{1}{\pi} Im \,\Pi(s,T)_{a} \;\theta(s - s_0)\;, \label{hadronicQ}
\end{equation}

where $s \equiv q^2 = \omega^2 - \mathbf{q}^2$, and the leptonic decay constant is defined as 

\begin{equation}
\langle 0| V_\mu(0) | V(k)\rangle = \sqrt{2}\; M_V \;f_V \;\epsilon_\mu \; . \label{FV}
\end{equation}

Next, considering a finite (total) width  the following replacement will be understood

\begin{equation}
\delta(s- M_V^2(T)) \Longrightarrow const \; \frac{1}{(s-M_V^2(T))^2 + M_V^2(T) \Gamma_V^2(T)}\; , \label{DeltaV}
\end{equation}

where the constant is fixed by requiring equality of areas, e.g. if the integration is in the interval $(0-infty)$, then $ const =  M_V(T) \Gamma_V(T)/\pi$. 

To complete the hadronic parametrization one needs  the hadronic scattering term due to the current scattering off  heavy-light quark pseudoscalar mesons (D-mesons). The expression in Eq.(\ref{Pi-V}) needs to be re-obtained, in principle,  as it is valid for massless pseudoscalar hadrons(pions). In the massive case it becomes

\begin{equation}
\frac{1}{\pi}\, Im \,\Pi^s(\omega,T)|_{HAD} =  \frac{2}{ 3 \pi^2}\; M_D^2\;\delta(\omega^2) \left[ n_B\left(\frac{M_D}{T}\right) + \frac{2 \,T^2}{M_D^2}\;\int_{m_D/T }^{\infty} y \,n_B(y)  \, dy \right] \;. \label{scattD}
\end{equation}

It is easy to verify that this term is exponentially suppressed, numerically being two to three orders of magnitude smaller than its QCD counterpart, Eq.(\ref{PIVQq0}).\\

Turning to the sum rules, the vector correlation function $\Pi(q^2,T)$, Eq.(\ref{VVcorrelator}), satisfies a once-subtracted dispersion relation. Hence, one can use Hilbert moments, Eqs.(\ref{duality})-(\ref{HilbertQCD}). The non-perturbative QCD term, of dimension $d=4$, corresponding to the gluon condensate is given by

\begin{eqnarray}
\varphi_N(Q^2,T)|_{NP} &=& - \frac{3}{4\pi^2}\frac{1}{(4m_Q^2)^N}\frac{1}{(1+\xi)^{N+2}}\;F\left(N+2,\,-\frac{1}{2}\,,N+\frac{7}{2},\,\rho\right) \nonumber \\ [.3cm]
&\times& \frac{2^N \,N\,(N+1)^2\,(N+2)\,(N+3)\,(N-1)!}{(2N+5)\,(2N+3)!!}\;\Phi
\label{PhiNP}
\end{eqnarray}

where $F(a,b,c,z)$ is the hyper-geometric function, , $\xi\equiv\frac{Q^2}{4m_Q^2}$, $\rho\equiv\frac{\xi}{1+\xi}$, and

\begin{equation}
\Phi\equiv\frac{4\pi^2}{9}\frac{1}{(4m_Q^2)^2} \left<\frac{\alpha_s}{\pi}G^2\right>|_T\;.\label{GGT}
\end{equation}

The thermal behaviour of the gluon condensate, needed as an input, was obtained from LQCD results available at the time \cite{LQCDGGT1}-\cite{LQCDGGT2}. Those results are  in good agreement  with the most recent ones \cite{LQCDGG} shown in Fig.4. The first three Hilbert moments, and four ratios, were considered in \cite{ccbar1} to determine the thermal behaviour of the four quantities, $s_0(T)$, $M_V(T)$, $\Gamma_V(T)$, and $f_V(T)$. Details of the procedure are thoroughly discussed in \cite{ccbar1}, so we proceed to discuss the results. Figure 19 shows the behaviour of the normalized continuum threshold, $s_0(T)/s_0(0)$. Unlike the situation in the light-quark sector, where this ratio approaches zero quite rapidly close to $T_c$ (see Fig. 6), in the $J/\psi$ channel $s_0(T)$ shows a dramatically different behaviour. In fact, $s_0(T)$ 
decreases by only some 10\% at $T=T_c$, as shown in Fig.19. At $T \simeq 1.2 \, T_c$ the decrease is only close to 40\%. Above this temperature the sum rules no longer have solutions, as there is no support for the integrals in the Hilbert moments. This is something which happens generally, regardless of the type of current entering the correlation functions for light or heavy quarks. The unequivocal interpretation of this result is that the $J/\psi$ survives above the critical temperature for deconfinement. This puts to rest the historical expectation \cite{MS} of the melting of charmonium at, or close to $T=T_c$. Further evidence is provided by the behaviour of the width, Fig. 20. While initially the width behaves as in light, and heavy-light quark systems, by increasing with increasing $T$, just above $T_c$ the width has a sharp turnaround, decreasing substantially, thus suggesting survival of the $J/\psi$. Finally, the behaviour of the coupling, increasing (rather than decreasing) sharply with temperature, as shown in Fig. 21, provides an unambiguous evidence for the survival of this state. Contrary to the thermal behaviour of these quantities, the mass hardly changes with temperature, as shown in Fig. 22.\\
The thermal behaviour of these four parameters in the scalar and pseudoscalar charmonium \cite{ccbar2}, as well as in the vector and pseudoscalar bottonoium \cite{bbar}, are very similar to the $J/\psi$. 

\section{QCD phase diagram at finite $T$ and baryon chemical potential}

In this section we outline the extension of the analysis of the thermal axial-vector current correlator, Section 3, to finite baryon chemical potential \cite{mu}. The starting point is the light-quark axial-vector current correlator, Eq.(\ref{AAcorrelator}), and the two-point function $\Pi_0(q^2)$. In the static limit $(\bf{q} \rightarrow 0)$, to leading order in PQCD, for finite $T$ and quark chemical potential $\mu_q$, with $\mu_q = \mu_B/3$, the function $\Pi_0(q^2)$ now becomes $\Pi_0(\omega^2,T,\mu_q)$, and is given by

\begin{equation}
   \frac{1}{\pi}{\mbox{Im}}\Pi_0(s)|_{PQCD}
   =
   \frac{1}{4\pi^2}\left[1-\tilde{n}_+\left(\frac{\sqrt{s}}{2}\right)
   -\tilde{n}_-\left(\frac{\sqrt{s}}{2}\right)\right]  -\frac{2}{\pi^2} T^2 \delta (s)\left[
   {\mbox{Li}}_2(-e^{\mu_B/T})
   +{\mbox{Li}}_2(-e^{-\mu_B/T})\right],
\label{pertQCD}
\end{equation}

where ${\mbox{Li}}_2(x)$ is the dilogarithm function, $s=\omega^2$, and the Fermi-Dirac thermal distributions for particles (anti-particles) are given by

\begin{equation}
   \tilde{n}_\pm(x)=\frac{1}{e^{(x\mp \mu_q)/T}+1}\,.
\label{F-D}
\end{equation}

In the limit where $T$ and/or $\mu_B$ are large compared to a mass scale, e.g. the quark mass, Eq.~(\ref{pertQCD}) becomes

\begin{equation}
   \frac{1}{\pi}{\mbox{Im}}\Pi_0(s)|_{PQCD}=
   \frac{1}{4\pi^2} \left[1-\tilde{n}_+\left(\frac{\sqrt{s}}{2}\right)
  -\tilde{n}_-\left(\frac{\sqrt{s}}{2}\right) \right]
  +\frac{1}{\pi^2}\;\delta(s)
   \left(\mu_q^2 + \frac{\pi^2T^2}{3}\right)\;.
\label{pertQCDlargeTmu}
\end{equation}

The hadronic spectral function, Eq.(\ref{PionPole0}), is

\begin{equation}
   \frac{1}{\pi}{\mbox{Im}}\Pi (s)
   |_{HAD}
   = 2 \; f_\pi^2(T,\mu_q)\; \delta (s)\,.
\label{HAD}
\end{equation}

\begin{figure}[ht!] 
\begin{center}
\includegraphics[height=3.3 in,width=4.7 in]{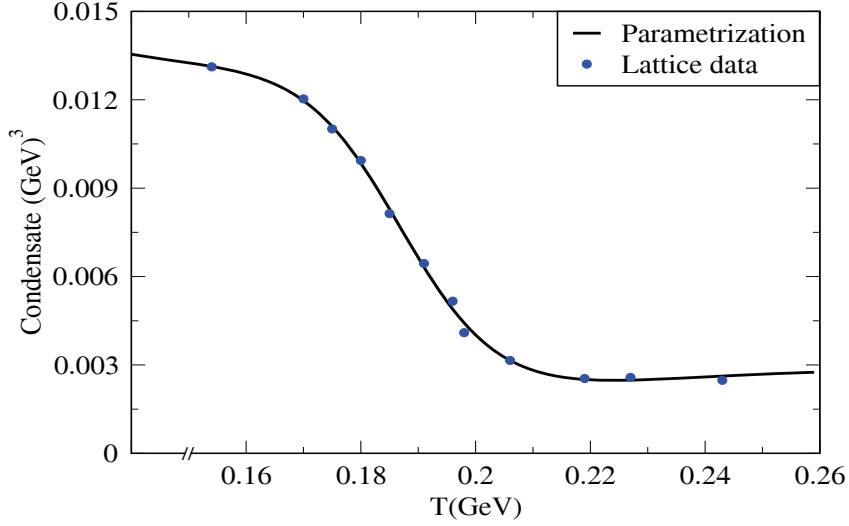}
\caption{\protect\small{LQCD data (dots) \cite{LQCDqq2}-\cite{LQCDqq3}, and absolute value of the quark condensate, $\langle \bar{\psi} \psi \rangle(T)$, Eq.(\ref{condexpl}) (solid curve), as a
  function of $T$ in the phase transition (or crossover) region.}} 
  \end{center}
\label{qbarqT}
\end{figure}

\begin{figure}[ht!]
\begin{center}
{\includegraphics [height=3.5 in,width=4.8 in] {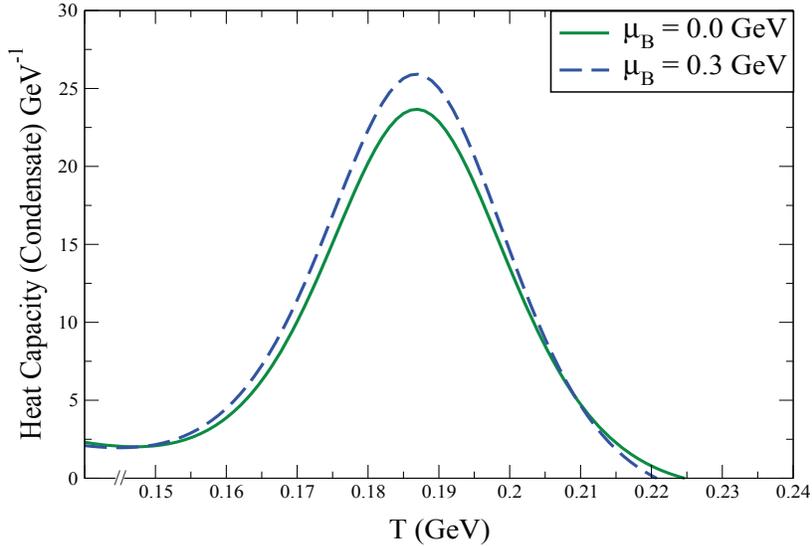}}
\caption{\protect\small{Heat capacity for the quark condensate as a function of $T$ for $\mu_B =0$ (solid line) and $\mu_B= 300 \; {\mbox{MeV}}$ (dash line). The critical temperature $T_c$ corresponds to the maximum of the heat capacity for a given value of $\mu_B$. }}  
\label{heat_capacity}
\end{center}
\end{figure}

Turning to the FESR, Eq.(\ref{FESR2}), with $N=1$, and using Eqs.(\ref{pertQCD}) and (\ref{HAD}) one finds

\begin{equation}
\int_0^{s_0(T,\mu_q)} ds \left[1 - \tilde{n}_+\left(\frac{\sqrt{s}}{2}\right) 
   -\tilde{n}_-\left(\frac{\sqrt{s}}{2}\right)\right] = 
   8\pi^2 f_\pi^2(T,\mu_q) + 8\, T^2\left[{\mbox{Li}}_2(-e^{\mu_q/T})
   + {\mbox{Li}}_2(-e^{-\mu_q/T})\right]. \label{FESRTMU}
\end{equation}

This transcendental equation determines $s_0(T,\mu_q)$ in terms of $f_\pi(T,\mu_q)$. The latter is related to the light-quark condensate through the Gell-Mann-Oakes-Renner relation \cite{GMORT}

\begin{equation}
\frac{f_\pi^2(T,\mu_q)}{f_\pi^2(0,0)} = \frac{\langle\bar{\psi}\psi\rangle(T,\mu_q)}{\langle\bar{\psi}\psi\rangle(0,0)}\;, \label{GMOR2}
\end{equation}

where the quark and pion masses were assumed independent of $T$ and $\mu_q$ in \cite{mu}.  In view of the results obtained in \cite{mqT}, as discussed in Section 6, it would seem important to revisit this issue. It is easy to see that a $T$-dependent quark mass does not affect the validity of Eq.(\ref{GMOR2}). In fact, the thermal quark mass follows the thermal pion mass, independently of $f_\pi(T)$ which in turn follows $\langle \bar{q} q\rangle (T)$. \\
A good closed form approximation to the FESR, Eq.(\ref{FESRTMU}), for large $T$ and/or $\mu_q$ is obtained using Eq.(\ref{pertQCDlargeTmu}) with $\tilde{n}_+\left(\frac{\sqrt{s}}{2}\right) \simeq \tilde{n}_-\left(\frac{\sqrt{s}}{2}\right) \simeq 0$, in which case

\begin{equation}
   s_0(T,\mu_q) \simeq  8 \, \pi^2\,
   f_\pi^2(T,\mu_q)
   -
   \frac{4}{3} \,\pi^2\, T^2 - 4 \, \mu_q^2\;.
\label{s0fpi}
\end{equation}

Using Eq.(\ref{GMOR2}) this can be rewritten as

\begin{equation}
  \frac{s_0(T,\mu_q)}{s_0(0,0)} \simeq \frac{\langle\bar{\psi}\psi\rangle (T,\mu_q)}{\langle\bar{\psi}\psi\rangle (0,0)}
   -
   \frac{(T^2/3 - \mu_q^2/\pi^2)}{2 f_\pi^2(0,0)}
\label{s0cond}
\end{equation}

The quark condensate can be computed from the in-medium quark propagator, whose non-perturbative properties can be obtained e.g. from known solutions to the Schwinger-Dyson equations (SDE) as discussed in detail in \cite{mu}. The result is

\begin{equation}
   \langle\bar{\psi}\psi\rangle(T,\mu_q)|_{\mbox{\tiny{matt}}} = - 
  \frac{8 T N_c}{\pi^2} \;
  \sum_{l=1}^\infty\frac{(-1)^l}{l}\cosh \left(\frac{\mu_q^l}{T} \right) \;
      \sum_{i=1}^4\frac{r_im_i^2}{|b_i|^3}\,
   K_1\left(\frac{l|m_i|}{T}\right),
\label{condexpl}
\end{equation}

where $K_1(x)$ is a Bessel function, and for convenience one defines $b_i=1$ for $\ i=1,2,3$, and $\ b_4=b$. The values of the parameters $m_i$, $r_i$, and $b_4\equiv b$ are given in Table I, and Table II in \cite{mu}. In the limit $\mu_B=0$, the result for the quark condensate using Eq.(\ref{condexpl}) is shown in Fig.23, together with LQCD data \cite{LQCDqq2}-\cite{LQCDqq3}
.\\
\begin{figure}[ht!] 
\begin{center}
\includegraphics[height=3.5 in,width=5.0 in]{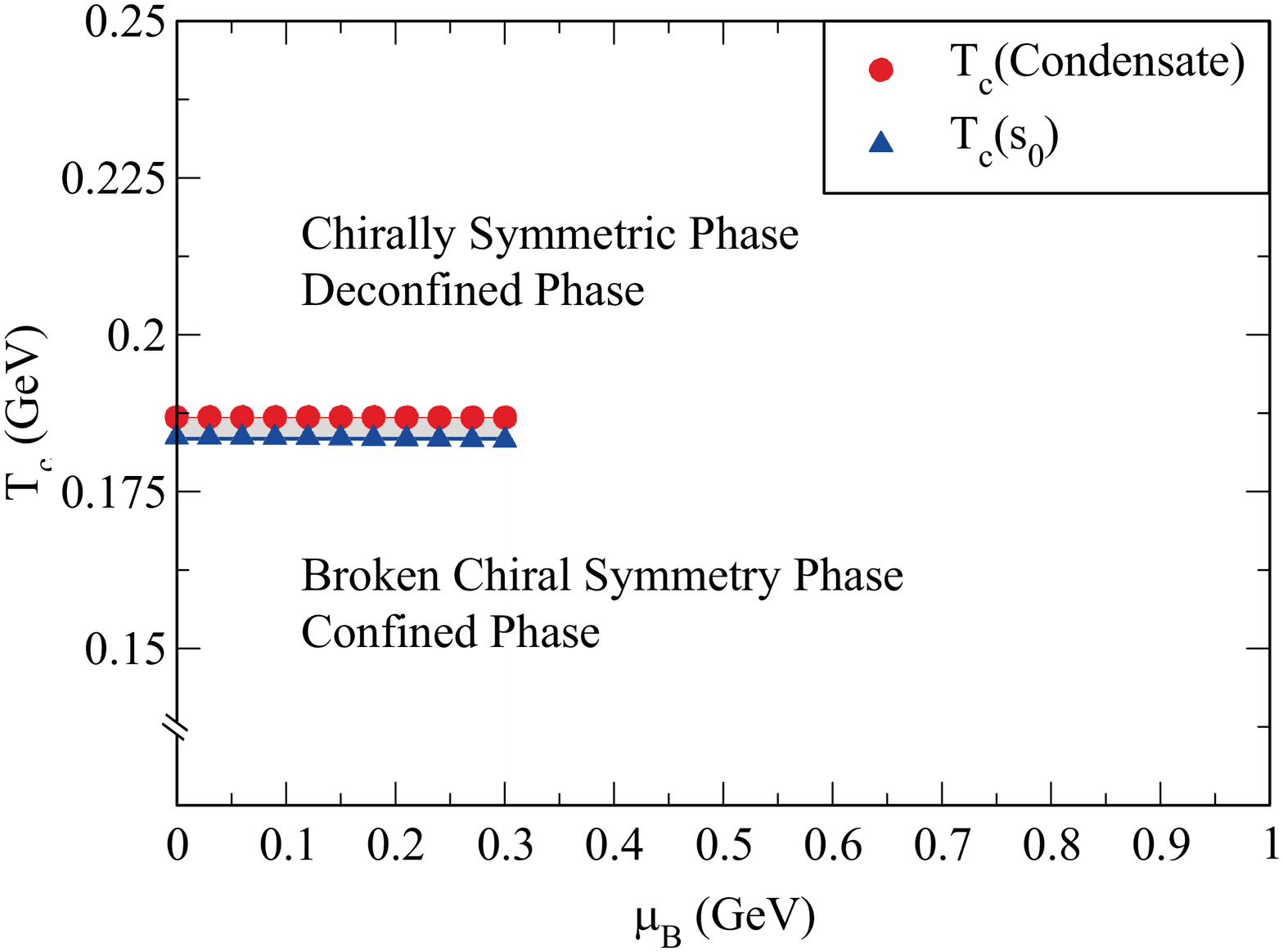}
\caption{\protect\small{Transition temperatures for the quark condensate, $\langle \bar{\psi} \psi\rangle(T, \mu_q)$ and the PQCD threshold, $s_0(T, \mu_q)$ as functions of the baryon chemical potential.}}  
  \end{center}
\label{phase}
\end{figure}

The expressions for $s_0(T,\mu_q)$ and $\langle \bar{\psi} \psi\rangle(T, \mu_q)$, Eqs. (\ref{s0cond}), (\ref{condexpl}), characterizing  deconfinement and  chiral-symmetry restoration transitions, are the central results of this analysis. They are used next to explore the phase diagram. To this end, one needs the corresponding susceptibilities, proportional to the heat capacities, $- \partial \langle \bar{\psi} \psi\rangle(T, \mu_q)/\partial T$, and $- \partial s_0 / \partial T$, for a given value of $\mu_B$. The transition temperature is then identified as the value of $T$ for which the heat capacity reaches a maximum. Figure 24 shows the behaviour of the heat capacity for the quark condensate (normalized to its value in the vacuum) as a function of $T$ for $\mu_B=0$ (solid line), and $\mu_B= 300 \, {\mbox{MeV}}$ (dash line). The  PQCD threshold, $s_0(T,\mu_B)$, is somewhat broader than the quark condensate (see \cite{mu}), but it peaks at essentially the same temperature, i.e. $T = 185\, \,{\mbox{MeV}}$, within $3 \,{\mbox{MeV}}$. The results for the phase diagram, $T_c$ versus $\mu_B$ are shown in Fig. 25, where the solid dots correspond to $T_c$ for chiral-symmetry restoration (quark condensate), and the solid triangles refer to deconfinement ($s_0$).

\section{Summary}

The extension of the QCD sum rule programme at $T=0$ \cite{Review1} to finite temperature was first proposed in \cite{BS} in the framework of Laplace transform QCDSR \cite{SVZ}. There are two main assumptions behind this extension, (i) the OPE of current correlators at short distances remains valid, except that the vacuum condensates will acquire a temperature dependence, and (ii) the notion of quark-hadron duality can be invoked in order to relate QCD to hadronic physics. The latter is known to be violated at $T=0$, in the low energy resonance region, DV, albeit by a relative small amount. This is unimportant at finite $T$, not only because of the small relative size of DV, but also because all determinations are normalized to their values at $T=0$. Next, the starting point is the identification of the basic object at finite $T$. This is the retarded (advanced) two-point function after appropriate Gibbs averaging

\begin{equation}
\Pi(q,T) = i \, \int \, d^4x \; e^{i q x}\; \theta(x_0)\; \langle \langle \left[ J(x), J^\dagger(0)\right] \rangle\rangle \,, \label{correlator2}
\end{equation}

where

\begin{equation}
\langle\langle  A \cdot B \rangle\rangle = \sum_n {\mbox{exp}}(- E_n/T) \; \langle n| A \cdot B| n\rangle \; / \; Tr \left( \mbox{exp} (-H/T)\right) \,, \label{Gibbs}
\end{equation}

and $|n\rangle$ is a {\bf complete} set of eigenstates of the (QCD) Hamiltonian. The OPE of $\Pi(q,T)$ is now written as

\begin{equation}
\Pi(q,T) = C_I \;\langle\langle I\rangle\rangle \;+\; C_r (q) \;\langle\langle {\cal{O}}_r\rangle\rangle \,.\label{OPET}
\end{equation}

It is essential to stress that the states $|n\rangle$ entering Eq.(\ref{Gibbs}) can be {\bf any} states, as long as they form a complete set. In other words, they could be hadronic states, or quark-gluon basis, etc.. The hadronic (mostly pionic) basis was advocated to obtain thermal information on some quantities, e.g. vacuum condensates \cite{EL95}. These determinations are constrained to very low temperatures, in the domain of thermal chiral perturbation theory, way below $T_c$. This approach does not invoke quark-hadron duality, thus it has little relationship to the QCD sum rule program. In addition, being restricted to very low temperatures, it provides no useful thermal information on e.g. QCD condensates, which is currently provided by LQCD. Alternatively, another complete set is the quark-gluon of QCD, as first advocated in \cite{BS}. This choice allows for a smooth extension of the QCDSR method to finite $T$. The only thermal restriction has to do with the support of the integrals entering the sum rules. In most cases this extends up to the critical temperature, an exception being charmonium which goes even further. Field theory arguments fully supporting this approach were given in \cite{QFTT}.\\

Another key element in this programme is the identification of the relevant QCD and hadronic parameters characterizing the transition to deconfinement and chiral-symmetry restoration. While the latter is universally understood to be the thermal quark condensate, an order parameter, in the case of deconfinement the parameter is purely phenomenological. It also differs from that used by LQCD, i.e. the so-called Polyakov loop. Thermal QCD sum rules invoke, instead, the onset of perturbative QCD in the square-energy, $s$-plane, so called $s_0(T)$, as first proposed in \cite{BS}. This choice is supported {\it a posteriori} by all applications in the light-quark and the heavy-light quark sector, resulting in an $s_0(T)$ decreasing monotonically with increasing temperature, and eventually vanishing at a critical temperature $T=T_c$. An important exception to this behaviour is the heavy-heavy quark system, i.e. charmonium (vector, scalar, and pseudoscalar channels) \cite{ccbar1}-\cite{ccbar2}, and bottonium \cite{bbar} (vector and pseudoscalar), for which $s_0(T)$ remains well above zero at or beyond $T_c$. Crucial theoretical validation of the role played by $s_0(T)$ has been obtained recently in \cite{Polyakov}, where a direct relation was found between $s_0(T)$ and LQCD's Polyakov thermal loop.

On the hadronic sector, the relevant parameters are the current-hadron coupling, and the hadronic width, both of which 
underpin the conclusions derived from the behaviour of $s_0(T)$, to wit. For light and heavy-light quark systems the current-hadron coupling decreases,  and the hadronic width increases monotonically with increasing $T$, thus signalling deconfinement. Instead, for the heavy-heavy quark systems, the coupling actually increases, and the width while initially growing, it reverses behaviour decreasing close to $T_c$, indicating the survival of these hadrons at and above $T_c$. This prediction was later confirmed for bottonium by LQCD \cite{LQCD_QQbar1}-\cite{LQCD_QQbar2}.\\
Another fundamental issue, to which this method contributed, was the relation between the two phase transitions, i.e. deconfinement and chiral-symmetry restoration. After preliminary indications of the approximate equality of both critical temperatures \cite{FESRT1}, a later analysis \cite{Gatto4} supported this conclusion. Recently, a more refined updated analysis \cite{FESRTAA} fully confirmed earlier results.\\
The extension of the well known Weinberg sum rules \cite{WSR0} to finite $T$, without prejudice on some pre-existing chiral mixing scenario \cite{Dey}, clearly show their full saturation, except very close to $T_c$, albeit returning to full saturation at $T=T_c$. These deviations are caused by the thermal space-like cut in the energy plane, arising at leading order in the vector channel, but loop suppressed in the axial-vector case. This asymmetric contribution, growing with the square of the temperature, vanishes at $T=T_c$. Hence, this feature has no relation whatsoever with a potential chiral mixing scenario. In fact, an inspection of the thermal behaviour of the hadronic parameters in the vector and the axial-vector channel fully disprove this idea. These spectral functions remain quite distinct at all temperatures, except at $T=T_c$ where they vanish for obvious reasons. In any case, and as shown in Section 5, as well as in \cite{Mallik}, in a hadronic thermal bath there is a chiral asymmetry due to Isospin and G-parity preventing any mixing.\\
On a separate issue, thermal QCD sum rules allow to determine the behaviour of the light quark masses, $m_{u,d}$ together with the pion decay constant, $f_\pi(T)$ \cite{mqT}. The two sum rules for the light-quark pseudoscalar axial-vector current divergence require as input the $T$-dependence of the pion mass \cite{MPIT}, and the quark condensate \cite{LQCD}. The result for $f_\pi(T)$ is fully consistent with chiral-symmetry, in that it follows the behaviour of $|\langle \bar{q}q\rangle(T)|$, independently of $M_\pi(T)$ (see Eqs. (\ref{Mpi2}), (\ref{fpi2})). It is also consistent with the expectation that close to $T_c$ the quark mass should increase, becoming the constituent mass at deconfinement.
Finally, QCDSR have been extended to finite $T$ together with finite baryon chemical potential, $\mu_B$, \cite{mu}. This has allowed to obtain the phase diagram ($T_c, \mu_B$). It should be possible, in future, to extend the explored range of $\mu_B$, as well as study other applications at finite $T$ and $\mu_B$.\\

A topic not discussed here is that of non-diagonal (Lorentz non-invariant) condensates. Clearly, the existence of a medium, i.e. the thermal bath, breaks trivially Lorentz invariance. However, after choosing a reference system at rest with respect to the medium, one can ignore this issue and continue to use a covariant formulation. Nevertheless, there might exist new terms in the OPE, absent at $T=0$. In the case of non-gluonic operators it has been shown that they are highly suppressed \cite{heavylight}, \cite{nondiag1} so that they can be ignored. A gluonic twist-two term in the OPE was considered in \cite{nondiag2}, and computed on the lattice in \cite{nondiag3}-\cite{nondiag4}. Once again, the contribution of such a term is negligible in comparison with all regular (diagonal) terms as shown in  \cite{ccbar1}.\\

In closing we wish to briefly mention a few applications of thermal QCDSR which were nor covered here. An independent validation of this method was obtained by determining the thermal behaviour of certain three-point functions (form factors), and in particular their associated root-mean-squared (rms) radii. In the case of the electromagnetic form factor of the pion, $F_\pi(q^2,T)$, it was found in \cite{pionffT} that it decreases with increasing $T$, almost independently of $Q^2$. The pion radius, $\langle r_\pi\rangle(T)$, increases with temperature, doubling at $T/T_c \simeq 0.8$, and diverging at $T\simeq T_c$, thus signalling deconfinement. On a separate issue, the axial-vector coupling of the nucleon, $g_A(T)$ was found to be essentially constant in most of the temperature range, except very close to $T_c$, where it starts to grow \cite{gAT}. The associated rms, $\langle r_A^2\rangle(T)$ was also found to be largely constant, but diverging close to $T_c$, consistent with deconfinement. This information was used to determine the thermal behaviour of the $SU(2) \times SU(2)$ Goldberger-Treiman relation (GTR), and its deviation, $\Delta_\pi$ defined in \cite{gAT} as

\begin{equation}
\frac{f_\pi(T) \, g_{\pi NN}(T)}{M_N(T) \, g_A(T)} \equiv 1 \, + \, \Delta_\pi(T) \,,
\label{GTRT}
\end{equation}

which is different from the standard definition $\Delta_\pi= 1 - M g_A/f_\pi g_{\pi NN}$. Given that the nucleon mass is basically independent of $T$, except very close to $T_c$ \cite{nucleonT}-\cite{nucleonT2}, and similarly for $g_A(T)$, the deviation $\Delta_\pi(T)$ decreases with increasing $T$ and the GTR ceases to be valid. \\
Another thermal three-point function analysis dealt with the coupling $g_{\rho \pi \pi}$, the associated rms radius, and the issue of the Vector Meson Dominance (VMD) at finite temperature \cite{VMDT}- \cite{Pisarski2}. Results from \cite{VMDT} indicated the approximate validity of an extension of VMD where the strong coupling $g_{\rho \pi \pi}$ becomes a function of the momentum transfer. This extended coupling decreases with increasing temperature, vanishing just before $T = T_c$, and the associated rms radius diverges close to the critical temperature, thus siganlling deconfinement.\\
Finally, the Adler-Bell-Jackiw axial anomaly \cite{ABJ1}-\cite{ABJ2} at finite $T$ was studied at low temperatures in \cite{Pisarski3}, and in the whole $T$ range in \cite{anomaly}. Results from \cite{Pisarski3} showed that the amplitude of $\pi^0 \rightarrow \gamma \, \gamma$ decreased with increasing T. The same behaviour was found in \cite{anomaly}, leading to the vanishing of that amplitude, provided VMD remains valid.

\begin{center}
{ \bf Acknowledgements}
\end{center}
The work of A.A. was supported in part by  UNAM-DGAPA-PAPIIT grant number IN101515, and by Consejo Nacional de Ciencia y Tecnologia grant number 256494.
The work of M.L. was supported in part by Fondecyt 1130056, Fondecyt 1150847 (Chile), and Proyecto Basal (Chile) FB 0821.
This work was also supported by NRF (South Africa), and the Research Administration, University of Cape Town.

\section{Appendix}
In this Appendix we derive the QCD expression of the QCD scattering term for a vector current correlation function of  non-zero (equal mass) quarks. Extensions to other currents and/or unequal quark masses should be straightforward. We begin with the correlator, Eq.(\ref{VVcorrelator}), in the time-like region. Substituting in Eq.(\ref{VVcorrelator}) the current $V_\mu(x) = : \overline{Q}^a(x) \, \gamma_\mu \, Q^a(x):$, where $Q(x)$ is a quark field of mass $m_Q$, and $a$ the colour index, results in

\begin{equation}
\Pi_{\mu\nu}^{a}(q^2) \equiv (-g_{\mu\nu}\, q^2 + q_\mu q_\nu) \, \Pi(q^2) = - i^3 \, N_c \int d^4 x \; e^{i q x} \; Tr \left[ \gamma_\mu \, S_F(x) \, \gamma_\nu\, S_F(-x) \right] 
\,, \label{PImunu}
\end{equation}

where $S_F(x)$ is the quark propagator in space-time, and $N_c=3$. Transforming the propagators to momentum-space, performing the integrations, and taking the imaginary part of $\Pi(q^2)$ gives

\begin{equation}
Im \;\Pi^{a}(q^2) = \frac{3}{16 \, \pi} \int_{-v}^{+v} dx \; (1- x^2) = \frac{1}{8 \, \pi} \,v (3 - v^2), \label{PIq}
\end{equation}

where the variable $v \equiv v(q^2)$ is given by

\begin{equation}
v(q^2) = \, \left(1 - \frac{4 \, m_Q^2}{q^2} \right)^{1/2}\,. \label{vq2}
\end{equation}
\begin{figure}[ht!] 
\begin{center}
\includegraphics[height=3.0 in,width=3.5 in]{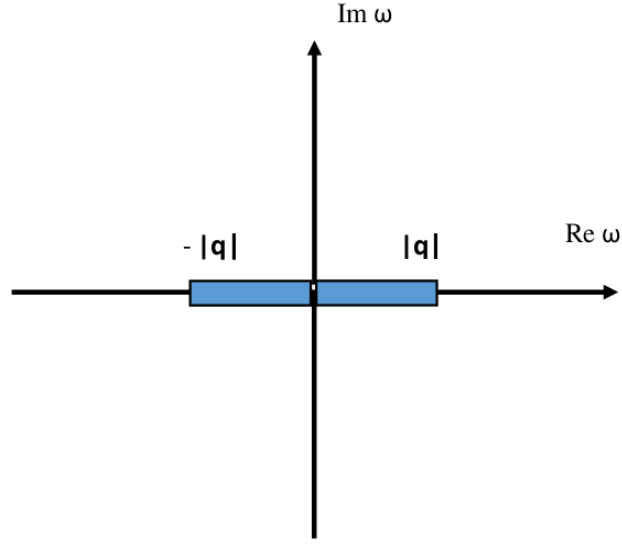}
\caption{\protect\small{The complex energy, $\omega$-plane, showing the central cut around the origin (scattering term), extending between $\omega = -|{\bf q}|$, and $\omega =|{\bf q}|$. The standard (time-like) annihilation right-hand and left-hand cuts at $\omega = \pm \left[|{\bf q}|^2 + \omega^2_{th}\right]^{1/2}$ are not shown ($\omega_{th}$ is some channel dependent mass threshold).}}
  \end{center}
\label{figvq2}
\end{figure}

Notice that because of the particular form of the current, in this case the normalization factor of $\Pi(q^2)$ for massless quarks is $Im \, \Pi(q^2) = 1/(4 \, \pi)$, instead of $1/(8 \, \pi)$ as in Eq.(\ref{Pi}).\\
The extension to finite $T$ can be performed using the Dolan-Jackiw thermal propagators, Eq.(\ref{DolanJackiw}), in Eq.(\ref{PImunu}), to obtain
.
\begin{equation}
Im \;\Pi^{a}(q^2, T) = \frac{3}{16 \, \pi} \int_{-v}^{+v} dx \; 
(1- x^2) \, \left[ 1 - n_F\left( \frac{|{\bf q}| x \, + \omega  }{2T}\right) - n_F\left( \frac{|{\bf q}| x \, - \omega  }{2T}\right)\right]\;.   \label{PIqT}
\end{equation}

In the rest frame of the medium, $|\bf q| \rightarrow 0$, this reduces to

\begin{equation}
Im \;\Pi^{a}(\omega, T) = \frac{3}{16 \, \pi} \int_{-v}^{+v} dx \; 
(1- x^2) \, \left[ 1 - 2 \, n_F\left( \omega / 2T \right) \right]\; = \frac{3}{16 \, \pi} \int_{-v}^{+v} dx \; 
(1- x^2) \, \mbox{tanh} \left( \frac{\omega}{4 T}\right)\;.
   \label{PIqT2}
\end{equation}

Proceeding to the scattering term, the equivalent to Eq.(\ref{PIqT}) is

\begin{equation}
Im \;\Pi^{s}(q^2, T) = \frac{3}{8 \, \pi} \int_{v}^{\infty} dx \; 
(1- x^2) \, \left[n_F\left( \frac{|{\bf q}| x \, + \omega  }{2T}\right) - n_F\left( \frac{|{\bf q}| x \, - \omega  }{2T}\right)\right]\;,  \label{PIqTs}
\end{equation}

where the integration limits arise from the bounds in the angular integration in momentum space. Notice that this term vanishes identically at $T=0$, and the overall multiplicative factor is twice the one in Eq.(\ref{PIqT}). Next, the thermal difference in the integrand can be converted into a derivative

\begin{equation}
Im \;\Pi^{s}(q^2, T) = \frac{3}{8\, \pi}\, \frac{\omega}{T}\; \int_{v}^{\infty} dx \; 
(1- x^2) \, \frac{d}{dy}\, n_F(y) \, \label{ImPis}
\end{equation}

where $y = |{\bf {q}}| \, x / (2 \,T)$. This expression reduces to

\begin{equation}
Im \;\Pi^{s}(q^2, T) =  \frac{3}{4 \, \pi} \, \frac{\omega}{|\bf {q}|} \, \left[ - n_F \left(\frac{|{\bf {q}}|\, v}{2 T}\right) (1 - v^2) + \frac{8 \,T^2}{|\bf {q}|^2} \;\int_{|{\bf {q}}|\, v / 2T} ^{\infty} \;y \; n_F(y) \, dy \right] \,.\label{ImPis2}
\end{equation}

In the limit $|{\bf q}|\rightarrow 0$ this result becomes

\begin{equation}
Im \;\Pi^{s}(q^2, T) =  \frac{3}{\pi} \, \lim_{\stackrel{|{\bf q}|\rightarrow 0}{\omega \rightarrow 0}} \;  \, \frac{\omega}{{\bf |q|}^3} \;m_Q^2 \left[ n_F \left(\frac{m_Q} {T}\right) + \frac{2 \,T^2}{m_Q^2} \;\int_{m_Q/T} ^{\infty} \;y \; n_F(y) \, dy \right] \,.\label{ImPis3}
\end{equation}

After careful performance of the limit, in the order indicated, the singular term $\omega/{\bf{|q|}}^3$ above becomes a delta function 

\begin{equation}
\lim_{\stackrel{|{\bf q}|\rightarrow 0}{\omega \rightarrow 0}} \;  \, \frac{\omega}{{\bf |q|}^3} = \frac{2}{3} \; \delta (\omega^2)\,, \label{deltalimit}
\end{equation}

and the final result for the scattering term is

\begin{equation}
Im \;\Pi^{s}(\omega, T) = \frac{2}{\pi} \, m_Q^2 \, \delta(\omega^2) \left[n_F \left( \frac{m_Q}{T} \right) + \frac{2 \, T^2}{m_Q^2} \,\int_{m_Q/T} ^{\infty} \;y \; n_F(y) \, dy \right] \,.\label{ImPis4} 
\end{equation}

Depending on the correlator, the limiting function, Eq.(\ref{deltalimit}), could instead be less singular in $|{\bf q}|$, in which case the scattering term vanishes identically.

\end{document}